\documentclass[english,aps,preprint]{revtex4}
\usepackage[latin9]{inputenc}
\usepackage{textcomp}
\usepackage{amsmath}
\usepackage{amssymb}
\usepackage{esint}

\makeatletter
\@ifundefined{textcolor}{}
{%
 \definecolor{BLACK}{gray}{0}
 \definecolor{WHITE}{gray}{1}
 \definecolor{RED}{rgb}{1,0,0}
 \definecolor{GREEN}{rgb}{0,1,0}
 \definecolor{BLUE}{rgb}{0,0,1}
 \definecolor{CYAN}{cmyk}{1,0,0,0}
 \definecolor{MAGENTA}{cmyk}{0,1,0,0}
 \definecolor{YELLOW}{cmyk}{0,0,1,0}
}

\makeatother

\usepackage{babel}
\begin{document}

\title{The modified Poynting theorem and the concept of mutual energy}

\author{Shuang-ren Zhao}

\email{}

\homepage{http://imrecons.com}

\selectlanguage{english}%

\affiliation{Imrecons Inc, Toronto, Canada }

\author{Kevin Yang}

\affiliation{Imrecons Inc, Toronto, Canada}

\author{Kang Yang}

\affiliation{Imrecons Inc, Toronto, Canada}

\author{Xingang Yang}

\affiliation{Imrecons Inc, Toronto, Canada}

\author{Xintie Yang}

\affiliation{Northwestern Polytechnical university, Xi'an, China}
\begin{abstract}
The goal of this article is to derive the reciprocity theorem, mutual
energy theorem from Poynting theorem instead of from Maxwell equation.
In this way the reciprocity theorem will become the energy theorem.
In order to realize this purpose the followings have been done. The
Poynting theorem is generalized to the modified Poynting theorem.
In the modified Poynting theorem the electromagnetic field is superimposition
of different electromagnetic fields including the field of retarded
potential and advanced potential, electric/magnetic mirrored field,
time-reversed field, time-offset field, space-offset field. The media
epsilon (permittivity) and mu (permeability) can also be different
in the different fields. The concept of mutual energy is introduced
which is the difference between the total energy and self-energy.
First we try to derive the mutual energy theorem from complex Poynting
theorem, it is failed. As a side effect we obtained the mixed mutual
energy theorem. We applied the average process to derive the mutual
energy theorem from Poynting theorem. This is derivation is not strictly.
Then we derive the mutual energy from Fourier domain, instead of obtained
the mutual energy theorem from time-domain. We obtain the time-reversed
mutual energy theorem. A time-reverse transform needed to further
derive the mutual energy theorem. The time-reverse transform contains
some information from Maxwell equation, hence the derivation is not
a purely derivation from Poynting theorem. Then we derive the mutual
energy theorem in time-domain. Using the modified Poynting theorem
with the concept of the mutual energy. The instantaneous modified
mutual energy theorem is derived. Applying time-offset transform and
time integral to the instantaneous modified mutual energy theorem,
the time-correlation modified mutual energy theorem is obtained. Assume
there are two electromagnetic fields one is retarded potential and
one is advanced potential, the convolution reciprocity theorem can
be derived. Corresponding to the modified time-correlation mutual
energy theorem and the time-convolution reciprocity theorem in Fourier
domain, there is the modified mutual energy theorem and the Lorentz
reciprocity theorem. Hence all mutual energy theorem and the reciprocity
theorems are put in one frame of the concept of the mutual energy.
The inner product is introduced for two different electromagnetic
fields in both time domain and Fourier domain. The concept of inner
product of electromagnetic fields simplifies the theory of the wave
expansion. The concept of reaction is re-explained as the mutual energy
of two fields with retarded potential and advanced potential.
\end{abstract}
\maketitle

\section{Introduction}

In electromagnetic field theory, the Poynting theorem\cite{Poynting}
is energy conservation theorem. The Lorentz reciprocity theorem\cite{Lorentz-1,Carson-1,Carson-2,STUART-BALLANTINE} 

\[
\intop_{S}(E_{1}(\omega)\times H_{2}(\omega)-E_{2}(\omega)\times H_{1}(\omega))\, dt\,\hat{n}dS
\]

\begin{equation}
=\intop_{V}\,(J_{1}(\omega)\cdot E_{2}(\omega)-J_{2}(\omega)\cdot E_{1}(\omega)-K_{1}(\omega)\cdot H_{2}(\omega)+K_{2}(\omega)\cdot H_{1}(\omega)\,\, dV=0\label{eq:10000-10}
\end{equation}
is close related to Poynting theorem. The two theorems look similar,
J.R. Carson call the reciprocity theorem as ``Reciprocal energy theorem''
in the reference\cite{Carson-2}. But until now the two theorems are
two different theorems derived from Maxwell equations respectively. 

Many efforts try to reveal the relationship between the reciprocity
theorem and Poynting theorem have been done. V. H. Rumsey proposed
the concept of the reaction\cite{Rumsey} in 1954 which is related
to Lorentz reciprocity theorem. But what is the concept of the ``reaction''
in behind scene?. W.J. Welch has derived a reciprocity theorem\cite{Welch}
in 1960, which is in the following,
\[
-\intop_{S}\intop_{t=-\infty}^{\infty}(E_{1}(t)\times H_{2}(t)+E_{2}(t)\times H_{1}(t))\, dt\,\hat{n}dS
\]
\begin{equation}
=\intop_{V}\intop_{t=-\infty}^{\infty}(J_{1}(t)\cdot E_{2}(t)+K_{1}(t)\cdot H_{2}(t)+J_{2}(t)\cdot E_{1}(t)+K_{2}(t)\cdot H_{1}(t))\, dt\, dV\label{eq:10000-20}
\end{equation}
In the formula all variables with subscript ``1'' is belong to retarded
potential and all variables with subscript ``2'' belong to advanced
potential. Welch has derived another reciprocity theorem in 1961,
\[
\intop_{S}\intop_{t=-\infty}^{\infty}(E_{1}(t)\times H_{2}(t)-E_{2}(t)\times H_{1}(t))\, dt\,\hat{n}dS
\]
\begin{equation}
=\intop_{V}\intop_{t=-\infty}^{\infty}(J_{1}(t)\cdot E_{2}(t)-K_{1}(t)\cdot H_{2}(t)-J_{2}(t)\cdot E_{1}(t)+K_{2}(t)\cdot H_{1}(t))\, dt\, dV\label{eq:10000-30}
\end{equation}
Rumsey has derived a another reciprocity theorem in 1963\cite{Rumsey_VH}
from his reaction concept, which is the following,

\begin{equation}
\intop_{V}(J_{1}(\omega)\cdot E_{2}^{*}(\omega)+J_{2}^{*}(\omega)\cdot E_{1}(\omega)+K_{1}(\omega)\cdot H_{2}^{*}(\omega)+K_{2}^{*}(\omega)\cdot H_{1}(\omega))\, dV=0\label{eq:10000-40}
\end{equation}
S. N. Samaddar suggest a reciprocity theorem and applied it to solve
wave expansion problems in 1964\cite{Samaddar} in plasma media,

\begin{equation}
\nabla\cdot(E_{l}\times\hat{H}_{l'}+\hat{E_{l'}}\times H_{l})+\hat{v}_{el'}\cdot p_{el}+v_{el}\hat{p}_{el'}+\hat{v}_{il'}\cdot p_{il}+v_{il}\hat{p}_{il'}++\hat{v}_{nl'}\cdot p_{nl}+v_{in]m}\hat{p}_{il'}=0\label{eq:10000-50}
\end{equation}
The corresponding time-domain theory of the reciprocity theorem\cite{Lorentz-1,Carson-1,Carson-2,STUART-BALLANTINE}\cite{Rumsey}
is the time-convolution reciprocity theorem\cite{Goubau,Ru-shao}
derived by G. Goubau in 1960 and B. Ru-shao Cheo in 1965 which is

\begin{equation}
\intop_{V}\,\intop_{-\infty}^{\infty}(J_{1}(\tau-t)\cdot E_{2}(t)-J_{2}(t)\cdot E_{1}(\tau-t)-K_{1}(\tau-t)\cdot H_{2}(t)+K_{2}(t)\cdot H_{1}(\tau-t)\, dt\, dV=0\label{eq:10000-60}
\end{equation}
The further development of the time-convolution reciprocity theorem
can be found\cite{Adrianus,Anders}. J. A. Kong offers the details
of conjugate transform and the concept of the modified reciprocity
theorem\cite{JinAuKong,J_A_Kong2} in 1972. In the modified reciprocity
theorem the $\epsilon$ (permittivity) and $\mu$ (permeability) of
two electromagnetic fields appeared in the reciprocity theorem are
allowed to be different. Norbert N. Bojarski has further developed
the Welch's reciprocity theorem in 1983\cite{NorberN}. Shuang-Ren
Zhao proposed the mutual energy theorem and modified mutual energy
theorem in May of 1987\cite{shrzhao1}, which is

\[
-\intop_{S}(E_{1}(\omega)\times H_{2}^{*}(\omega)+E_{2}^{*}(\omega)\times H_{1}(\omega))\cdot\hat{n}dS
\]
\begin{equation}
=\intop_{V}(J_{1}(\omega)\cdot E_{2}^{*}(\omega)+J_{2}^{*}(\omega)\cdot E_{1}(\omega)+K_{1}(\omega)\cdot H_{2}^{*}(\omega)+K_{2}^{*}(\omega)\cdot H_{1}(\omega))\, dV\label{eq:10000-80}
\end{equation}
The derivation of the mutual energy theorem is based on Lorenz reciprocity
theorem\cite{Lorentz-1,Carson-1,Carson-2,STUART-BALLANTINE}\cite{Rumsey}\cite{J_A_Kong2}
and the conjugate transform\cite{J_A_Kong2}. The mutual energy theorem
is defined in Fourier domain or complex domain and is further developed
in the reference\cite{shrzhao2,shrzhao3}. Compare to the Rumsey's
formula in the mutual energy theorem there is an item of the surface
integral which dose not vanish. The surface integral has been applied
to define an inner product of electromagnetic fields on the surface
and hence to solve the wave expansion problems. Welch's reciprocity
theorem\cite{Welch,NorberN} is further developed by A. T. de Hoop
in December 1987 to become the so called time-correlation reciprocity
theorem\cite{Adrianus2} which is
\[
-\intop_{S}\intop_{t=-\infty}^{\infty}(E_{1}(t+\tau)\times H_{2}(t)+E_{2}(t)\times H_{1}(t+\tau))\, dt\,\hat{n}dS
\]
\begin{equation}
=\intop_{V}\intop_{t=-\infty}^{\infty}(J_{1}(t+\tau)\cdot E_{2}(t)+K_{1}(t+\tau)\cdot H_{2}(t)+J_{2}(t)\cdot E_{1}(t+\tau)+K_{2}(t)\cdot H_{1}(t+\tau))\, dt\, dV\label{eq:10000-90}
\end{equation}
de Hoop's time-correlation reciprocity theorem can be seen as the
mutual energy theorem\cite{shrzhao1} in time-domain instead of in
the Fourier domain. Welch's reciprocity theorem is a special situation
of the time-correlation reciprocity theorem where the time variable
$\tau=0$. By the way in the theorem of de Hoop\cite{Adrianus2} the
surface integral appeared but has been thought that it will vanish
on the infinite sphere. Baun has a book\cite{Baum-1} in 1995 which
systematically introduced reciprocity theorems. The reference \cite{D_Marcuse,A_A_Barybin}
solved the wave expansion problem directly from Maxwell equation which
is close related to the mutual energy theorem. 

Later the mutual energy theorem has been rediscovered and has been
referred as the second reciprocity theorem\cite{Petrusenko} in 2009.
In this second reciprocity theorem the surface integral was also thought
to be vanish on the infinite big sphere. 

Application of reciprocity theorem and mutual energy theorem can also
be found in the examples\cite{Steven_G_Johnson,Anatoly_A_Barybin,T_D_Carozzi,JiupingChen}.
There are a few reference discussed the relationship between Poynting
theorem and reciprocity theorem\cite{SHUN-LIEN_CHUANG,Burak_Polat,Wen_Cho_Chew}.
However they only discussed them together and did not offer the direct
relationship between the two theorems. The reference \cite{JinAuKong}
discussed the reciprocity theorem in bi-anisotropic media. 

There is a concept ``mixed Poynting vector''\cite{Fragstein-Conrad,Angus-Macleod}
which is close related to the concept mutual energy and Poynting vector.

It is worth to notice that in the reference\cite{shrzhao1,shrzhao2,shrzhao3},
the concept of ``mutual energy'' was not well defined and the related
concept ``total energy'' and ``self energy'' were also not defined.
The so called mutual energy theorem is only derived from the modified
reciprocity theorem instead of Poynting theorem, hence the concept
of mutual energy is still not widely acceptable.

\section{The contribution of this article}

The goal of this article is to convince the reader that the mutual
energy theorem is a real energy. To achieve this goal the mutual energy
theorem have to be derived from Poynting theorem instead of from Maxwell
equation directly. In this article a few new concept is defined or
generalized. Among them there is ``modified'', ``time-reversed
transform'', ``mutual energy'', ``self energy'', ``total energy''.
The concept of reaction is reexplained. The mutual energy theorem
is re-derived from Poynting theorem. The derivation include different
versions and the history is introduced. A few new theorem is obtained
from the derivation, the following gives the details.

\subsection{The modified Maxwell equation}

The concept of ``modified'' is borrowed from the modified reciprocity
theorem, where two electromagnetic field put in different media with
different $\epsilon$ and $\mu$ can be superimposed. We found for
this situation the Maxwell equation is still established. In this
kind of media, the Maxwell equation is referred as modified Maxwell
equation.

\subsection{The modified Poynting theorem}

The Poynting theorem is generalized to the modified Poynting theorem.
The concept of ``modified'' is borrowed from the modified reciprocity
theorem\cite{J_A_Kong2} as above modified Maxwell equation. This
idea has also been used in the mutual energy theorem which is modified
mutual energy\cite{shrzhao1,shrzhao2,shrzhao3}.

\subsection{Time-reversed transform}

We knew that after a magnetic mirror transform the electromagnetic
field $(E,H)$ is still electromagnetic field. A electromagnetic field
after a time-reversed transform is not a electromagnetic field any
more. That means after the time-reversed transform it does not satisfy
the Maxwell equation. However we modified the time-reversed transform
through introducing the negative media with negative $\epsilon$ ,
$\mu$, this new time-reversed transform is given in the following,

\[
[E_{r}(t),H_{r}(t),J_{r}(t),K_{r}(t),\epsilon_{r}(t),\mu_{r}(t)]
\]
\[
\equiv r[E(t),H(t),J(t),K(t),\epsilon(t),\mu(t)]
\]
\begin{equation}
=[E(-t),H(-t),J(-t),K(-t),-\epsilon(-t),-\mu(-t)]\label{eq:2000-500}
\end{equation}
A electromagnetic fields after the above time-reverse transform is
still electromagnetic fields. This time reverse transform is one of
the import tool to derive mutual energy theorem from Poynting theorem
in Fourier domain.

\subsection{The difference the substitution and the replace of a transform is
noticed}

There are many transforms: magnetic mirror transform, electric mirror
transform, time-reversed transform. Time-offset transform. There are
two process for the above transform, one is substitution of the transform,
the other is replace of a transform. Replace and substitution are
two different process, in the history many mistakes were made since
the confusion with this two processes. In this article we try to clarify
the difference of the replacement and substitution of a transform.

\subsection{Introduced the instantaneous mutual energy theorem}

In this article the concept of mutual energy is defined as the difference
between the total energy and the self energy. The instantaneous mutual
energy theorem is derived from Poynting theorem with the concept of
mutual energy. The instantaneous mutual energy theorem is following,

\[
-\nabla\cdot(E_{1}\times H_{2}+E_{2}\times H_{1})
\]

\begin{equation}
=J_{1}\cdot E_{2}+J_{2}\cdot E_{1}+K_{1}\cdot H_{2}+K_{2}\cdot H_{1}+E_{1}\cdot\partial D_{2}+E_{2}\cdot\partial D_{1}+H_{1}\cdot\partial B_{2}+H_{2}\cdot\partial B_{1}\label{eq:90-1}
\end{equation}

\subsection{Derived the time-reversed mutual energy theorem in Fourier domain}

Time-reversed mutual energy theorem is introduced from Fourier domain
through instantaneous mutual energy theorem, and hence from the Poynting
theorem. The reversed mutual energy theorem in Fourier domain is show
in the following,
\[
-\intop_{S}(E_{1}(\omega)\times H_{2}(\omega)+E_{2}(\omega)\times H_{1}(\omega))\:\hat{n}dS
\]

\begin{equation}
=\intop_{V}(J_{1}(\omega)\cdot E_{2}(\omega)+J_{2}(\omega)\cdot E_{1}(\omega)+K_{1}(\omega)\cdot H_{2}(\omega)+K_{2}(\omega)\cdot H_{1}(\omega))\, dV\label{eq:7200-250-1}
\end{equation}
The time-reversed mutual energy theorem in time domain is

\[
-\intop_{S}\intop_{t=-\infty}^{\infty}(E_{1}(\tau-t)\times H_{2}(t)+E_{2}(t)\times H_{1}(\tau-t))\:\hat{n}dS
\]
\begin{equation}
=\intop_{V}\intop_{t=-\infty}^{\infty}(J_{1}(\tau-t)\cdot E_{2}(t)+J_{2}(t)\cdot E_{1}(\tau-t)+K_{1}(\tau-t)\cdot H_{2}(t)+K_{2}(t)\cdot H_{1}(\tau-t))\, dV\label{eq:7300-50-1}
\end{equation}

\subsection{Derived the mutual energy theorem in Fourier domain}

The mutual energy theorem can be derived from reversed mutual energy
theorem with time-reversed transform. The mutual energy theorem is
shown in the following,

\[
-\intop_{S}(E_{1}(\omega)\times H_{2}^{*}(\omega)+E_{2}^{*}(\omega)\times H_{1}(\omega))\cdot\hat{n}dS
\]
\begin{equation}
=\intop_{V}(J_{1}(\omega)\cdot E_{2}^{*}(\omega)+J_{2}^{*}(\omega)\cdot E_{1}(\omega)+K_{1}(\omega)\cdot H_{2}^{*}(\omega)+K_{2}^{*}(\omega)\cdot H_{1}(\omega))\, dV\label{eq:10000-80-1}
\end{equation}
The corresponding in the time domain which is time-correlation mutual
energy theorem or time-correlation reciprocity theorem\cite{Adrianus2}.

\[
-\intop_{S}\intop_{t=-\infty}^{\infty}(E_{1}(t+\tau)\times H_{2}(t)+E_{2}(t)\times H_{1}(t+\tau))\, dt\,\hat{n}dS
\]
\begin{equation}
=\intop_{V}\intop_{t=-\infty}^{\infty}(J_{1}(t+\tau)\cdot E_{2}(t)+K_{1}(t+\tau)\cdot H_{2}(t)+J_{2}(t)\cdot E_{1}(t+\tau)+K_{2}(t)\cdot H_{1}(t+\tau))\, dt\, dV\label{eq:10000-90-1}
\end{equation}

\subsection{Derived the time-reversed reciprocity theorems from the mutual energy
theorems in Fourier domain}

The time-reversed reciprocity theorem is shown as following,
\[
\intop_{S}(E_{1}(\omega)\times H_{2}^{*}(\omega)-E_{2}(\omega)\times H_{1}^{*}(\omega))\, dt\,\hat{n}dS
\]

\begin{equation}
=\intop_{V}\,(J_{1}(\omega)\cdot E_{2}^{*}(\omega)-J_{2}(\omega)\cdot E_{1}^{*}(\omega)-K_{1}(\omega)\cdot H_{2}^{*}(\omega)+K_{2}^{*}(\omega)\cdot H_{1}(\omega)\,\, dV=0\label{eq:10000-100}
\end{equation}
In time-domain, the corresponding theorem is time-correlation reciprocity
theorem is derived which is in the following,

\[
\intop_{S}\intop_{t=-\infty}^{\infty}(E_{1}(t)\times H_{2}(t+\tau)-E_{2}(t+\tau)\times H_{1}(t))\, dt\,\hat{n}dS
\]
\begin{equation}
=\intop_{V}\intop_{t=-\infty}^{\infty}(J_{1}(t)\cdot E_{2}(t+\tau)-K_{1}(t)\cdot H_{2}(t+\tau)-J_{2}(t+\tau)\cdot E_{1}(t)+K_{2}(t+\tau)\cdot H_{1}(t))\, dt\, dV\label{eq:10000-110}
\end{equation}

\subsection{Re-derived the Lorenz reciprocity theorem from the mutual energy
theorem}

The Lorenz reciprocity theorem\cite{Lorentz-1}\cite{Rumsey} is shown
as a special situation of the mutual energy theorem. In the special
situation where two electromagnetic fields are different, one is the
field of retarded potential and the other is the field of advanced
potential. The concept of the reaction\cite{Rumsey} is re-explained
as the mutual energy (or power) of the two electromagnetic fields
where one is the field of retarded potential and the other is the
field of advanced potential.

\subsection{Re-derived time-correlation mutual energy theorem from Poynting theorem}

In the above we have derived the mutual energy theorem in Fourier
domain. The derivation is that first derive the time-reversed mutual
energy theorem and then through a time-reverse transform we obtained
the mutual energy theorem. We did not very satisfy with this process
of derivation. Since it is not a pure derivation from Poynting theorem.
The time-reversed transform need the Maxwell equation to prove. Hence
we actually derived the mutual energy theorem form Poynting theorem
plus Maxwell equation. A purely derivation from Poynting theorem should
not use the transform for example mirrored transform or time-reversed
transform. Hence we seek another way to prove the mutual energy theorem
and avoid the time-reversed transform. We have proved that the time-correlation
mutual energy theorem from Poynting theorem. And then considered a
Fourier transform for the time-correlation mutual energy theorem,
we obtained the mutual energy theorem. This way we have purely derived
the mutual energy theorem form the Poynting theorem.

\subsection{Introduced the mixed mutual energy theorem}

We have try to derive the mutual energy theorem from complex Poynting
theorem. But we have failed. Instead to obtained the mutual energy
theorem we obtained the following mixed mutual energy theorem. 

\[
-\intop_{S}(E_{1}\times H_{2}^{*}+E_{2}\times H_{1}^{*})\cdot\hat{n}dS
\]
\[
=\intop_{V}(E_{1}\cdot J_{2}^{*}+E_{2}\cdot J_{1}^{*}+H_{1}^{*}\cdot K_{2}+H_{2}^{*}\cdot K_{1})dV
\]
\begin{equation}
+j\omega\intop_{V}(H_{1}^{*}\cdot\mu_{2}H_{2}+H_{2}^{*}\cdot\mu_{1}H_{1}-E_{1}\cdot\epsilon_{2}^{*}E_{2}^{*}-E_{2}\cdot\epsilon_{1}^{*}E_{1}^{*})dV\label{eq:2000-200-1}
\end{equation}
or corresponding mixed time-correlation mutual energy theorem, 

\[
-\intop_{S}\intop_{t=-\infty}^{\infty}(E_{1}(t+\tau)\times H_{2}^{*}(t)+E_{2}(t+\tau)\times H_{1}^{*}(\tau))dt\cdot\hat{n}dS
\]
\[
=\intop_{V}\intop_{t=-\infty}^{\infty}(E_{1}(t+\tau)\cdot J_{2}^{*}(t)+E_{2}(t+\tau)\cdot J_{1}^{*}(t)+H_{1}^{*}(t)\cdot K_{2}(t+\tau)+H_{2}^{*}(t)\cdot K_{1}(t+\tau))\, dt\, dV
\]
\[
+\partial_{\tau}\intop_{V}\intop_{t=-\infty}^{\infty}(H_{1}^{*}(t)\cdot(\mu_{2}*H_{2})(t+\tau)+H_{2}^{*}(t)\cdot(\mu_{1}*H_{1})(t+\tau)
\]
\begin{equation}
-E_{1}(t+\tau)\cdot(\epsilon_{2}^{*}*E_{2}^{*})(t)-E_{2}(t+\tau)\cdot(\epsilon_{1}^{*}*E_{1}^{*})(t))\, dt\, dV\label{eq:2000-220}
\end{equation}
The mixed mutual energy theorem is related with the concept of mixed
Poynting vector \cite{Fragstein-Conrad,Angus-Macleod}. We do not
clear perhaps the mixed mutual energy has some usage in the future.

\subsection{Introduced the inner product to the mutual energy theorem}

This author has introduced the inner product in Fourier domain\cite{shrzhao1,shrzhao2,shrzhao3}.
In this article this idea is generalized to time domain. In time domain
the inner product is defined as following

\[
(\zeta_{1},\zeta_{2})_{\tau}=\intop_{S}\intop_{t=-\infty}^{\infty}(E_{1}(t+\tau)\times H_{2}(t)+E_{2}(t)\times H_{1}(t+\tau))\, dt\,\hat{n}dS
\]
The author has shown that $(\zeta_{1},\zeta_{2})_{\tau}$ is not a
good inner product, but $(\zeta_{1},\zeta_{2})_{\tau=0}$ is a good
inner product. The inner product is applied to the wave expansion
problem. The normal function expansion method can be applied to electromagnetic
field expansions.

\subsection{Re-explained the concept of the reaction}

Many confused concept about transform is clarified. There are time-reversed
transform, mirror transform, time offset transform. In the above transform
there are two different process in derivation of new theory one is
substitution and the other is replacement. Mistake is often caused
by confusing the replacement as substitution. The concept of causal
field, advanced potential, retard potential, offset field, transmitting
filed, receiving field is clarified too. The concept reaction is re-explained
as the mutual energy of two fields one is retarded potential and the
other one is advanced potential.

\subsection{Complementary theorems}

Chen-To Tai has derived the complementary reciprocity theorem\cite{Chen-to_tai}.
We have obtained 4 theorems, 2 mutual energy theorem and 2 reciprocity
theorem. We apply the electromagnetic field swapping transform 

\[
\zeta_{s}=s\zeta=[ZH,\frac{1}{Z}E,-\frac{1}{Z}K,-ZJ,-\frac{1}{Z^{2}}\mu,-Z^{2}\epsilon]
\]
4 corresponding complementary theorems are obtained. Among them one
is the Chen-To tai's complementary reciprocity theorem.

\section{Modified Maxwell equation}

\subsection{The Maxwell equation and the modified Maxwell equation }

There are two kinds of electromagnetic fields, one is transmitting
field and the other one is receiving field. Assume $\xi=[E,\, H]$
is radiated from the source $\rho=[J,K]$. $\rho$ is inside the volume
$V$, the boundary of the volume is $S=\partial V$. The example of
this kind of electromagnetic field is the electromagnetic field radiated
from an antenna. $\xi=[E,\, H]$ is the retarded potential. Another
kind of field is the field received by the sink of $\rho=[J,K]$.
The example of this kind of electromagnetic field is the electromagnetic
field received by an antenna which is advanced potential.

Assume $\zeta=[E,H,J,K,\epsilon,\mu]$ is a electromagnetic system,
where $\xi=[E,\, H]$ are electric field intensity and magnetic field
intensity. $\rho=[J,K]$ are electric current distribution and magnetic
current distribution. $\epsilon,\mu$ are permittivity and permeability,
we assume $\zeta$ satisfies the Maxwell equation, 

\begin{equation}
\nabla\times H=J+\partial D\label{eq:490-1}
\end{equation}
\begin{equation}
\nabla\times E=-K-\partial B\label{eq:500-1}
\end{equation}
where $\partial=\partial_{t}=\frac{\partial}{\partial t}$, $t$ is
time. $\nabla$ is gradient operator to the space variable $x=[x_{1},x_{2},x_{3}]$
is the rectangle coordinates, ``$\times$'' is vector cross product
operator. ``$\nabla\times$'' is ``$curl$'' operator. $\nabla\cdot$
is divergence operator Here $D$ is electric displacement field intensity;
$B$ is magnetic induction field intensity. And 
\begin{equation}
D(t)=\intop_{\tau=-\infty}^{\infty}\epsilon(t-\tau)E(\tau)\, d\tau\label{eq:502-1-2-1-1-0}
\end{equation}

\begin{equation}
B(t)=\intop_{\tau-\infty}^{\infty}\mu(t-\tau)H(\tau)d\tau\label{eq:502-1-2-1-1-1}
\end{equation}
If there is only one media $\epsilon,\mu$, the electromagnetic field
can also be written as $\zeta=[E,H,J,K,D,B]$. In general we assume
$\epsilon$ and $\mu$ are tensors
\begin{equation}
\epsilon=[\epsilon_{ij}],\ \ \ \ \ \ \mu=[\mu_{ij}]\ \ \ \ \ i=j=1,2,3\label{eq:3000-10}
\end{equation}

If there are $N$ electromagnetic fields $\zeta_{i}=[E_{i},H_{i},J_{i},K_{i}D_{i},B_{i}]$,
$i=1,2,...N$. Assume $\zeta_{1}$, $\zeta_{2}$ ... $\zeta_{N}$
satisfy the above Maxwell equation. Assume the superimposing electromagnetic
field is 
\begin{equation}
\zeta=\zeta_{1}+\zeta_{2}\cdots\zeta_{N}\label{eq:1000-01}
\end{equation}
There is the relationship,

\begin{equation}
D(t)=\intop_{\tau=-\infty}^{\infty}\epsilon(t-\tau)(E_{1}(\tau)+E_{2}(\tau)+\cdots+E_{N}(\tau))d\tau\label{eq:501-1-2-1-3}
\end{equation}
\begin{equation}
B(t)=\intop_{\tau=-\infty}^{\infty}\epsilon(t-\tau)(H_{1}(\tau)+H_{2}(\tau)+\cdots+H_{N}(\tau))d\tau\label{eq:502-1-2-1-3}
\end{equation}

\subsection{The modified Maxwell equation }

Where $\zeta=[E,H,J,K,D,B]$. In the space with media (permittivity
and permeability) $\epsilon,\mu$, normally $D$ and $B$ should satisfy
the above formula. For the above formula the $D(t)$ and $B(t)$ are
not linear. $D(t)$ and $B(t)$ are only linear when all fields $\zeta_{1}+\zeta_{2}\cdots\zeta_{N}$
has same media, i.e. 
\begin{equation}
\epsilon_{i=1,\cdots N}=\epsilon,\ \ \ \ \ \ \mu_{i=1,\cdots N}=\mu\label{eq:2100-499}
\end{equation}
The above relationship can be loosen by defining that the relation
from $D$ to $E$ and $B$ to $H$ are linear, which is.

\begin{equation}
D(t)=D_{1}(t)+D_{2}(t)+\cdots+D_{N}(t)\label{eq:2100-500}
\end{equation}
\begin{equation}
B(t)=B_{1}(t)+B_{2}(t)+\cdots+B_{N}(t)\label{eq:2100-510}
\end{equation}
where

\begin{equation}
D_{i}(t)=\intop_{\tau=-\infty}^{\infty}\epsilon_{i}(t-\tau)E_{i}(\tau)d\tau\ \ \ \ \ \ i=1,\cdots,N\label{eq:2100-520}
\end{equation}
\begin{equation}
B_{i}(t)=\intop_{\tau=-\infty}^{\infty}\mu_{i}(t-\tau)E_{i}(\tau)d\tau\ \ \ \ \ \ i=1,\cdots,N\label{eq:2100-530}
\end{equation}
Hence there is
\begin{equation}
D(t)=\intop_{\tau-\infty}^{\infty}(\epsilon_{1}(t-\tau)E_{1}(\tau)+\epsilon_{2}(t-\tau)E_{2}(\tau)+\cdots+\epsilon_{N}(t-\tau)E_{N}(\tau))d\tau\label{eq:501-1-2-1-1}
\end{equation}
\begin{equation}
B(t)=\intop_{\tau-\infty}^{\infty}(\mu_{1}(t-\tau)H_{1}(\tau)+\mu_{2}(t-\tau)H_{2}(\tau)+\cdots+\mu_{N}(t-\tau)H_{N}(\tau))d\tau\label{eq:502-1-2-1-1}
\end{equation}
This modification can also be found in reference\cite{Anders}. It
can be proven that if $\zeta_{1},\zeta_{2}\cdots\zeta_{N}$ satisfy
the Maxwell equation, with different media Eq(\ref{eq:2100-520},\ref{eq:2100-530}),
the superimposing electromagnetic field $\zeta=\zeta_{1}+\zeta_{2}+\cdots+\zeta_{N}$
will also satisfies Maxwell equation Eq.(\ref{eq:490-1},\ref{eq:500-1})
with the above loosen relation Eq.(\ref{eq:501-1-2-1-1},\ref{eq:502-1-2-1-1}).
In the following we will combine the media equation to Maxwell equation.
In case the field $\zeta$ satisfies the Maxwell equation with the
media Eq.(\ref{eq:501-1-2-1-3},\ref{eq:502-1-2-1-3}), it is will
be referred satisfying the Maxwell equation. In case the field satisfies
the Maxwell equation with the media condition Eq.(\ref{eq:501-1-2-1-1},\ref{eq:502-1-2-1-1})
it is referred as the modified Maxwell equation. The concept of ``modified''
is borrowed from the modified reciprocity theorem\cite{JinAuKong,J_A_Kong2}.

\section{The transform of electromagnetic filed}

\subsection{Time reverse transform}

Assume $r$ is time reversed transform\cite{Altman}, $\zeta=[E(t),H(t),J(t),K(t),\epsilon(t),\mu(t)]$
is electromagnetic system and satisfies Maxwell Equation. $[E_{r},H_{r},J_{r},K_{r},\epsilon_{r},\mu_{r}]$
is the transformed electromagnetic system. $\zeta_{r}=r\zeta$, or

\[
[E_{r}(t),H_{r}(t),J_{r}(t),K_{r}(t),D_{r}(t),B_{r}(t)]\equiv r[E(t),H(t),J(t),K(t),D(t),B(t)]
\]
\begin{equation}
=[E(-t),H(-t),J(-t),K(-t),D(-t),B(-t)]\label{eq:1000-20}
\end{equation}
It can be proved that if the electromagnetic field $\zeta$ satisfied
Maxwell equation, the time-reversed electromagnetic field $\zeta_{r}$
is also satisfies the the time reversed Maxwell equation: 

\begin{equation}
\nabla\times H=J-\partial D\label{eq:490-1-1}
\end{equation}
\begin{equation}
\nabla\times E=-K+\partial B\label{eq:500-1-1}
\end{equation}
Proof: If the electromagnetic field $\zeta$ is normal field (satisfies
Maxwell equation), If $\zeta$ is time reversed field then $r\zeta$
is normal field. $\zeta_{r}=r\zeta$ will be time-reversed satisfies
time-reversed Maxwell equation. There is $\zeta=r\zeta_{r}$ or

\[
[E(t),H(t),J(t),K(t),D(t),B(t)]\equiv r[E_{r}(t),H_{r}(t),J_{r}(t),K_{r}(t),D_{r}(t),B_{r}(t)]
\]
\begin{equation}
=[E_{r}(-t),H_{r}(-t),J_{r}(-t),K_{r}(-t),D_{r}(-t),B_{r}(-t)]\label{eq:1000-20-2}
\end{equation}
Substituting this to the Maxwell equation, there is

\begin{equation}
\nabla\times H_{r}(-t)=J_{r}(-t)+\partial_{t}D_{r}(-t)\label{eq:490-1-1-1}
\end{equation}
\begin{equation}
\nabla\times E_{r}(-t)=-K_{r}(-t)-\partial_{t}B_{r}(-t)\label{eq:500-1-1-1}
\end{equation}
Assume $-t=\tau$, $\partial_{t}=-\partial_{\text{\ensuremath{\tau}}}$
\begin{equation}
\nabla\times H_{r}(\tau)=J_{r}(\tau)-\partial_{\tau}D_{r}(\tau)\label{eq:490-1-1-1-1}
\end{equation}
\begin{equation}
\nabla\times E_{r}(\tau)=-K_{r}(\tau)+\partial{}_{\tau}B_{r}(\tau)\label{eq:500-1-1-1-1}
\end{equation}
Hence $\zeta_{r}$ satisfies the time reversed Maxwell equation. Proof
finish.

The above time reversed transform has been applied to produce a few
reciprocity theorems\cite{Samaddar}\cite{NorberN}. It is worth to
say, even the time-reverse transformed field does not satisfy the
Maxwell equation, but if we put the minus sign insider the media,
i.e., define

\begin{equation}
\epsilon_{r}(-t)=-\epsilon(-t),\ \ \ \ \ \ \ \mu_{r}(-t)=-\mu(-t)\label{eq:4100-100}
\end{equation}
or
\begin{equation}
[E_{r}(t),H_{r}(t),J_{r}(t),K_{r}(t),\epsilon_{r}(t),\mu_{r}(t)]\equiv r[E(t),H(t),J(t),K(t),\epsilon(t),\mu(t)]\label{eq:4100-110}
\end{equation}
\begin{equation}
=[E(-t),H(-t),J(-t),K(-t),-\epsilon(-t),-\mu(-t)]\label{eq:1000-20-4}
\end{equation}
We can prove that $\zeta_{r}=[E_{r}(t),H_{r}(t),J_{r}(t),K_{r}(t),\epsilon_{r}(t),\mu_{r}(t)]$
satisfies the Maxwell equation as following 
\begin{equation}
\nabla\times H_{r}(\tau)=J_{r}(\tau)+\partial_{\tau}(\epsilon_{r}(t)*E_{r}(\tau))\label{eq:490-1-1-1-1-1}
\end{equation}
\begin{equation}
\nabla\times E_{r}(\tau)=-K_{r}(\tau)-\partial{}_{\tau}(\mu_{r}(t)*H_{r}(\tau))\label{eq:500-1-1-1-1-1}
\end{equation}
Hence whether or not the time-reverse transformed field satisfies
Maxwell equation, is depending how the media after the transform is
defined. In the following only the Eq(\ref{eq:1000-20-4}) will be
referred as time reversed transform. A electromagnetic field after
time-reverse transform is still magnetic field.

\subsection{Magnetic mirror transform}

Assume $h$ is magnetic mirror transform\cite{Altman}, $\zeta=[E(t),H(t),J(t),K(t),\epsilon(t),\mu(t)]$
is electromagnetic field and satisfies Maxwell Equation. $[E_{h},H_{h},J_{h},K_{h},\epsilon_{h},\mu_{h}]$
is the transformed electromagnetic system. $\zeta_{h}=h\zeta$, or

\[
[E_{h}(t),H_{h}(t),J_{h}(t),K_{h}(t),\epsilon_{h}(t),\mu_{h}(t)]\equiv h[E(t),H(t),J(t),K(t),\epsilon(t),\mu(t)]
\]
\begin{equation}
=[E(-t),-H(-t),-J(-t),K(-t),\epsilon(-t),\mu(-t)]\label{eq:1000-20-1}
\end{equation}

$\zeta_{h}=h\zeta$, it can be easily proven that $\zeta=h\zeta_{h}$,
i.e.,

\[
[E(t),H(t),J(t),K(t),\epsilon(t),\mu(t)]=h[E_{h}(t),H_{h}(t),J_{h}(t),K_{h}(t),\epsilon_{h}(t),\mu_{h}(t)]
\]
\begin{equation}
=[E_{h}(-t),-H_{h}(-t),-J_{h}(-t),K_{h}(-t),\epsilon_{h}(-t),\mu_{h}(-t)]\label{eq:1000-20-1-2}
\end{equation}
It can be proved that if the electromagnetic field $\zeta$ satisfied
Maxwell equation, the magnetic mirror transformed field $\zeta_{h}$
also satisfies the Maxwell equation. 

Proof: Substitute Eq.(\ref{eq:1000-20-1-2}) to the Maxwell equation
Eq.(\ref{eq:490-1},\ref{eq:500-1})

\begin{equation}
\nabla\times(-1)H_{h}(-t)=(-1)J_{h}(-t)+\partial D_{h}(-t)\label{eq:490-1-2}
\end{equation}
\begin{equation}
\nabla\times E_{h}(-t)=-K_{h}-\partial(-1)B(-t)\label{eq:500-1-2}
\end{equation}
or

\begin{equation}
\nabla\times(-1)H_{h}(-t)=(-1)J_{h}(-t)+(-1)\partial_{-t}D_{h}(-t)\label{eq:490-1-2-1}
\end{equation}
\begin{equation}
\nabla\times E_{h}(-t)=-K_{h}-(-1)\partial_{-t}(-1)B(-t)\label{eq:500-1-2-1}
\end{equation}
or

\begin{equation}
\nabla\times H_{h}(-t)=J_{h}(-t)+\partial_{-t}D_{h}(-t)\label{eq:490-1-2-1-1}
\end{equation}
\begin{equation}
\nabla\times E_{h}(-t)=-K_{h}-\partial_{-t}B(-t)\label{eq:500-1-2-1-1}
\end{equation}
substitute $\tau=-t$, there is
\begin{equation}
\nabla\times H_{h}(\tau)=J_{h}(\tau)+\partial_{\tau}D_{h}(\tau)\label{eq:490-1-2-1-1-1}
\end{equation}
\begin{equation}
\nabla\times E_{h}(\tau)=-K_{h}-\partial_{\tau}B_{h}(\tau)\label{eq:500-1-2-1-1-1}
\end{equation}
Hence $\zeta_{h}$ satisfied the Maxwell equation. Proof finish.

Hence an electromagnetic fields after the magnetic mirror transform,
it is stall electromagnetic field satisfying Maxwell equation.

\subsection{Electric mirror transform}

Assume $e$ is electric reversed transform, $\zeta=[E(t),H(t),J(t),K(t),\epsilon(t),\mu(t)]$
is electromagnetic system and satisfies Maxwell Equation. $[E_{e},H_{e},J_{e},K_{e},\epsilon_{e},\mu_{e}]$
is the transformed electromagnetic system. $\zeta_{e}=e\zeta$, or

\[
[E_{e}(t),H_{e}(t),J_{e}(t),K_{e}(t),\epsilon_{e}(t),\mu_{e}(t)]\equiv e[E(t),H(t),J(t),K(t),\epsilon(t),\mu(t)]
\]
\begin{equation}
=[-E(-t),H(-t),J(-t),-K(-t),\epsilon(-t),\mu(-t)]\label{eq:1000-20-1-1}
\end{equation}
It can be proven that $\zeta_{e}=e\zeta$ satisfies Maxwell equation.

Proof: $\zeta=e\zeta_{e}$, substitute this to Maxwell equation Eq(\ref{eq:490-1},\ref{eq:500-1}).
There is

\begin{equation}
\nabla\times H_{e}(-t)=J_{e}(-t)+\partial(-1)D_{e}(-t)\label{eq:490-1-2-2}
\end{equation}
\begin{equation}
\nabla\times(-1)E_{e}(-t)=-(-1)K_{e}-\partial B_{e}(-t)\label{eq:500-1-2-2}
\end{equation}
or considering $\partial_{t}=(-1)\partial_{-t}$

\begin{equation}
\nabla\times H_{e}(-t)=J_{e}(-t)+(-1)\partial_{-t}(-1)D_{e}(-t)\label{eq:490-1-2-1-2}
\end{equation}
\begin{equation}
\nabla\times(-1)E_{e}(-t)=-(-1)K_{e}-(-1)\partial_{-t}B_{e}(-t)\label{eq:500-1-2-1-2}
\end{equation}
or

\begin{equation}
\nabla\times H_{e}(-t)=J_{e}(-t)+\partial_{-t}D_{e}(-t)\label{eq:490-1-2-1-1-2}
\end{equation}
\begin{equation}
\nabla\times E_{e}(-t)=-K_{e}-\partial_{-t}B(-t)\label{eq:500-1-2-1-1-2}
\end{equation}
or considering $-t=\tau$

\begin{equation}
\nabla\times H_{e}(\tau)=J_{e}(\tau)+\partial_{\tau}D_{e}(\tau)\label{eq:490-1-2-1-1-1-1}
\end{equation}
\begin{equation}
\nabla\times E_{e}(\tau)=-K_{e}-\partial_{\tau}B_{e}(\tau)\label{eq:500-1-2-1-1-1-1}
\end{equation}
Hence $\zeta_{e}$ is also satisfies the Maxwell equation too. Proof
finish.

\subsection{Conjugate transform corresponding to time-reversed transform}

Considering a real function $f(t)$,

\[
f(t)=F^{-1}\{F(\omega)\}\equiv\intop_{-\infty}^{\infty}F(\omega)\exp(j\omega t)\, dt
\]
Hence
\[
f(-t)=\intop_{-\infty}^{\infty}F(\omega)\exp(-j\omega t)\, dt
\]
\[
=\{\intop_{-\infty}^{\infty}F^{*}(\omega)\exp(j\omega t)\, dt\}^{*}
\]
Considering $f(-t)$ is a real function
\[
f(-t)=f^{*}(-t)
\]

\[
=\intop_{-\infty}^{\infty}F^{*}(\omega)\exp(j\omega t)\, dt
\]
\[
=F^{-1}\{F^{*}(\omega)\}
\]

In the time-reversed transform considering $f(-t)\rightarrow F^{*}(\omega)$,
we can obtained the corresponding time-reversed transform in Fourier
domain.

Assume $r$ is time reversed transform\cite{Altman}, $\zeta=[E(\omega),H(\omega),J(\omega),K(\omega),\epsilon(\omega),\mu(\omega)]$
is electromagnetic system and satisfies Maxwell Equation. $[E_{r},H_{r},J_{r},K_{r},\epsilon_{r},\mu_{r}]$
is the transformed electromagnetic system. $\zeta_{r}=r\zeta$, or

\[
[E_{r}(\omega),H_{r}(\omega),J_{r}(\omega),K_{r}(\omega),\epsilon_{r}(\omega),\mu_{r}(\omega)]\equiv r[E(\omega),H(\omega),J(\omega),K(\omega),\epsilon(\omega),\mu(\omega)]
\]
\begin{equation}
=[E^{*}(\omega),H^{*}(\omega),J^{*}(\omega),K^{*}(\omega),-\epsilon^{*}(\omega),-\mu^{*}(\omega)]\label{eq:1000-20-3}
\end{equation}
or 
\[
\zeta_{r}\equiv r\zeta
\]
There is

\[
\zeta=r\zeta_{r}
\]
or
\[
[E(\omega),H(\omega),J(\omega),K(\omega),\epsilon(\omega),\mu(\omega)]\equiv r[E_{r}(\omega),H_{r}(\omega),J_{r}(\omega),K_{r}(\omega),\epsilon_{r}(\omega),\mu_{r}(\omega)]
\]
\begin{equation}
=[E_{r}^{*}(\omega),H_{r}^{*}(\omega),J_{r}^{*}(\omega),K_{r}^{*}(\omega),-\epsilon_{r}^{*}(\omega),-\mu_{r}^{*}(\omega)]\label{eq:1000-20-3-1}
\end{equation}

Assume $\zeta=[E(\omega),H(\omega),J(\omega),K(\omega),D(\omega),B(\omega)]$
satisfies Maxwell Equation Eq.(Eq(\ref{eq:490-1},\ref{eq:500-1}).
).

\begin{equation}
\nabla\times H_{r}^{*}=J^{*}+(j\omega)(-\epsilon_{r}^{*})E_{r}^{*}\label{eq:490-1-1-2}
\end{equation}
\begin{equation}
\nabla\times E_{r}^{*}=-K_{r}^{*}-(j\omega)(-\mu_{r}^{*})H_{r}^{*}\label{eq:500-1-1-2}
\end{equation}
or
\begin{equation}
\nabla\times H_{r}=J_{r}+j\omega\epsilon_{r}E_{r}\label{eq:490-1-1-2-1}
\end{equation}
\begin{equation}
\nabla\times E_{r}=-K_{r}-j\omega\mu_{r}H_{r}\label{eq:500-1-1-2-1}
\end{equation}
Hence $[E_{r},H_{r}]$ is the solution of Maxwell equation with the
current $[J_{r},K_{r}]$ and media $[\epsilon_{r},\mu_{r}]$

\subsection{The conjugate transform corresponding magnetic mirror transform}

Considering $F\{f(-t)\}=f^{*}(\omega)$ which $F\{\bullet\}$ is Fourier
transform, tn Fourier domain (or complex space), the magnetic mirrored
transform become conjugate transform\cite{J_A_Kong2,JinAuKong}. Assume
$h$ is magnetic mirrored transform\cite{Altman}, $\zeta=[E(t),H(t),J(t),K(t),\epsilon(t),\mu(t)]$
is electromagnetic system and satisfies Maxwell Equation. $\zeta_{h}(t)=[E_{h}(t),H_{h}(t),J_{h}(t),K_{h}(t),\epsilon_{h}(t),\mu_{h}(t)]$
is the mirror transformed electromagnetic field, The corresponding
conjugate transform is defined as following $\zeta_{h}(\omega)=h\zeta(\omega)$,
or

\[
[E_{h}(\omega),H_{h}(\omega),J_{h}(\omega),K_{h}(\omega),\epsilon_{h}(\omega),\mu_{h}(\omega)]\equiv h[E(\omega),H(\omega),J(\omega),K(\omega),\epsilon(\omega),\mu(\omega)]
\]
\begin{equation}
=[E^{*}(\omega),-H^{*}(\omega),-J^{*}(\omega),K^{*}(\omega),\epsilon^{*}(\omega),\mu^{*}(\omega)]\label{eq:1000-20-1-3}
\end{equation}
Since $\zeta_{h}=h\zeta$, it can be easily proven that there is $\zeta=h\zeta_{h}$,
or

\[
[E(\omega),H(\omega),J(\omega),K(\omega),\epsilon(\omega),\mu(\omega)]=h[E_{h}(\omega),H_{h}(\omega),J_{h}(\omega),K_{h}(\omega),\epsilon_{h}(\omega),\mu_{h}(\omega)]
\]
\begin{equation}
=[E_{h}^{*}(\omega),-H_{h}^{*}(\omega),-J_{h}^{*}(\omega),K_{h}^{*}(\omega),\epsilon_{h}^{*}(\omega),\mu_{h}^{*}(\omega)]\label{eq:1000-20-1-2-1}
\end{equation}
If $\zeta(\omega)$ is electromagnetic field, after conjugate transform,
$h\zeta$ is also electromagnetic field.

\subsection{The difference between the replacement and the substitution of a
transform}

Assume there is a formula $f(...,\zeta)$ which contents the field
$\zeta=[J,K,E,H,\epsilon,\mu]$.

\begin{equation}
f(...,\zeta)=f(...,[J,K,E,H,\epsilon,\mu])\label{eq:4600-100}
\end{equation}

Here $f(...,\zeta)$ can be any known electromagnetic formula derived
from Maxwell equation for example the reciprocity theorem. 

There is difference between the replacement and substitution of a
transform. Assume here $[J,K]$ are the source. Assume $\zeta_{h}$
is the magnetic mirror transformed filed, so that $\zeta_{h}=[J_{h},K_{h},E_{h},H_{h},\epsilon_{h},\mu_{h}]=h\zeta$.
And hence there is $\zeta=h\zeta_{h}$. $h$ is mirror transform Hence
$\zeta=h\zeta_{h}=[-J_{h}(-t),K_{h}(-t),-H_{h}(-t),E(-t),\epsilon_{h}(-t),\mu_{h}(-t)]$.
we can substitute $\zeta=h\zeta_{h}$ to above formula, which is

\begin{equation}
f(...,\zeta)=f(...,[-J_{h}(-t),K_{h}(-t),E_{h}(-t),-H_{h}(-t),\epsilon_{h}(-t),\mu_{h}(-t)])\label{eq:4600-110}
\end{equation}
In this situation the formula does not change. However if we replace
the $\zeta$ using $\zeta_{h}$ , which is

\begin{equation}
f(...,\zeta_{h})=f(...,[J_{h},K_{h},E_{h},H_{h},\epsilon_{h},\mu_{h}])\label{eq:4600-120}
\end{equation}
The formula is changed. Since clearly that $\zeta$ is not $\zeta_{h}$.
Substituting $\zeta_{h}=h\zeta$ to the above formula, there is

\begin{equation}
f(...,\zeta_{h})=f(...,[-J(-t),K(-t),E(-t),-H(-t),\epsilon(-t),\mu(-t)])\label{eq:4600-130}
\end{equation}
 is a different formula compare to,

\[
f(...,\zeta)=f(...,[J,K,E,H,\epsilon,\mu])=
\]
\begin{equation}
f(...,[-J_{h}(-t),K_{h}(-t),E_{h}(-t),-H_{h}(-t),\epsilon_{h}(-t),\mu_{h}(-t)])\label{eq:4600-140}
\end{equation}

Hence if $f(...,\zeta)=0$ we can not guarantee that $f(...,\zeta_{h})=0$.

Substitution will not change the original formula, but the replacement
will change the original formula. Using replacement actually derive
a new formula which is the dual of the original formula. It mast be
very careful to the replacement and substitution of the transform
which are two different manipulations.

\subsection{Time offset transform }

Assume $T$ is time offset transform $\zeta=[E(t),H(t),J(t),K(t),\epsilon(t),\mu(t)]$
is electromagnetic system and satisfies Maxwell Equation. $\zeta_{T}=[E_{T},H_{T},J_{T},K_{T},\epsilon_{T},\mu_{T}]$
is the transformed electromagnetic system.
\[
\zeta_{T}\equiv[E_{T}(t),H_{T}(t),J_{T}(t),K_{T}(t),\epsilon_{T}(t),\mu_{T}(t)]
\]

\[
\equiv T[E(t),H(t),J(t),K(t),\epsilon(t),\mu(t)]
\]
\begin{equation}
=[E(t+T),H(t+T),J(t+T),K(t+T),\epsilon(t+T),\mu(t+T)]\label{eq:1000-30}
\end{equation}
It can be proved that if the electromagnetic field $\zeta$ satisfies
the Maxwell equation, then the time-offset electromagnetic field $\zeta_{T}$
is also satisfies the Maxwell equation.

\subsection{Space offset transform }

Assume $T$ is time offset transform $\zeta=[E(t),H(t),J(t),K(t),\epsilon(t),\mu(t)]$
is electromagnetic system and satisfies Maxwell Equation. $\zeta_{X}=[E_{X},H_{X},J_{X},K_{X},\epsilon_{X},\mu_{X}]$
is the transformed electromagnetic system. Where $X$ is stand for
space transform. $X$ is also a value of space offset $X=[X_{1},X_{2},X_{3}]$.
Assume $x=[x_{1},x_{2},x_{3}]$. 
\[
\zeta_{X}\equiv[E_{X}(t,x),H_{X}(t,x),J_{X}(t,x),K_{X}(t,x),\epsilon_{X}(t,x),\mu_{X}(t,x)]
\]

\[
\equiv X[E(t,x),H(t,x),J(t,x),K(t,x),\epsilon(t,x),\mu(t,x)]
\]
\begin{equation}
=[E(t,x+X),H(t,x+X),J(t,x+X),K(t,x+X),\epsilon(t,x+X),\mu(t,x+X)]\label{eq:1000-30-1}
\end{equation}
It can be proved that if the electromagnetic field $\zeta$ satisfies
the Maxwell equation, then the space-offset electromagnetic field
$\zeta_{X}$ is also satisfies the Maxwell equation. 

The idea of time offset and space offset has been used in reference\cite{NorberN}.

\subsection{Transform by swapping electric field and magnetic field}

Assume $s$ is a swap transform which can swap electric field and
magnetic field, $\zeta=[E,H,J,K,\epsilon,\mu]$
\[
\zeta_{s}=s\zeta=[aH,bE,cK,dJ,e\mu,f\epsilon]
\]
and this transform should be able to reverse, that means
\[
\zeta=s\zeta_{s}
\]
\[
\zeta=s\zeta_{s}=[aH_{s},bE_{s},cK_{s},dJ_{s},e\mu_{s},f\epsilon_{s}]
\]
Where $a,b,c,d,e,f$ are constant to be found. Assume $\zeta=[E,H,J,K,\epsilon,\mu]$
satisfy Maxwell equation
\[
\nabla\times H=J+j\omega\epsilon E
\]
\[
\nabla\times E=-K-j\omega\mu H
\]
We can substitute $\zeta$ to the above Maxwell equation, $\zeta_{s}=[E_{s},H_{s},J_{s},K_{s},\epsilon_{s},\mu_{s}]$
should also satisfy the Maxwell equation 

\[
\nabla\times H_{s}=J_{s}+j\omega\epsilon_{s}E_{s}
\]
\[
\nabla\times E_{s}=-K_{s}-j\omega\mu_{s}H_{s}
\]

After substituting we have
\[
\nabla\times(bE_{s})=cK_{s}+j\omega e\mu_{s}aH_{s}
\]
\[
\nabla\times(aH_{s})=-dJ_{s}-j\omega f\epsilon bE_{s}
\]
or

\[
\nabla\times E_{s}=\frac{c}{b}K_{s}+\frac{ea}{b}j\omega\mu_{s}H_{s}
\]
\[
\nabla\times H_{s}=-\frac{d}{a}J_{s}-\frac{fb}{a}j\omega\epsilon_{s}E_{s}
\]
We found that the constant should following 
\[
\frac{c}{b}=-1
\]
\[
-\frac{d}{a}=1
\]
\[
\frac{ea}{b}=-1
\]
\[
-\frac{fb}{a}=1
\]
Or
\[
c=-b
\]
\[
d=-a
\]
\[
e=-\frac{b}{a}
\]
\[
f=-\frac{a}{b}
\]
What about $a$ and $b$? Since this transform is not only satisfy
the Maxwell equation, there has unit dimension problem. After the
transform it should has same unit dimension. 
\[
E_{s}=aH
\]
\[
H=bE_{s}
\]
\[
E_{s}=abE_{s}
\]
or

\[
ab=1
\]

Now if $a$ is known all other constant can be found. $a$ actually
can be any value, however considering

\[
\epsilon_{0}E^{2},\ \ \ \mu_{0}H^{2}
\]
have the same unit dimension of energy, hence the following has the
same unit dimension. Here $\epsilon_{0}$ and $\mu_{0}$ is values
in empty space. 
\[
\sqrt{\epsilon_{0}}E,\ \ \ \sqrt{\mu_{0}}H
\]
or have same unit dimension.
\[
E,\ \ \ ZH
\]
where 
\[
Z=\sqrt{\frac{\mu_{0}}{\epsilon_{0}}}
\]
Hence we can just take 
\[
a=Z
\]
From above we can find if we take 
\[
b=\frac{1}{a}=\frac{1}{Z}
\]
Further we have

\[
c=-\frac{1}{Z}
\]
\[
d=-Z
\]
\[
e=-\frac{1}{Z^{2}}
\]
\[
f=-Z^{2}
\]
or

\[
\zeta_{s}=s\zeta=[ZH,\frac{1}{Z}E,-\frac{1}{Z}K,-ZJ,-\frac{1}{Z^{2}}\mu,-Z^{2}\epsilon]
\]
substitute $\zeta_{s}$ to
\[
\nabla\times H_{s}=J_{s}+j\omega\epsilon_{s}E_{s}
\]
\[
\nabla\times E_{s}=-K_{s}-j\omega\mu_{s}H_{s}
\]
we have

\[
\nabla\times(\frac{1}{Z}E)=(-\frac{1}{Z}K)+j\omega(-\frac{1}{Z^{2}}\mu)(ZH)
\]
\[
\nabla\times(ZH)=-(-ZJ)-j\omega(-Z^{2}\epsilon)(\frac{1}{Z}E)
\]
or
\[
\nabla\times E=-K-j\omega\mu H
\]
\[
\nabla\times H=J+j\omega\epsilon E
\]
That is the Maxwell equation.

\section{Advanced potential and retarded potential}

\subsection{Mirrored transform for $A(t),\phi(t),\varrho(t)$}

Assume $A(t),\phi(t)$ are vector potential and scale potential and
is defined as following 
\begin{equation}
E=-\nabla\phi-\partial A\label{eq:2100-10}
\end{equation}
\begin{equation}
B=\nabla\times A\label{eq:2100-20}
\end{equation}
Assume the magnetic field $\zeta=[E(t),H(t),J(t),K(t),\epsilon(t),\mu(t)]$.
Assume the magnetic current is assumed as $K=0$. Assume $\varrho$
is electric charge distribution which is related current distribution
through continue equation
\begin{equation}
\nabla\cdot J+\partial\varrho=0\label{eq:2100-60}
\end{equation}
Hence, $J$ is the only source of the potential $[A(t),\phi(t)]$.
Now let us find out the magnetic mirror transformed potential and
electric charge distribution $[A(t),\phi(t),\varrho(t)]$, i.e. $[A_{h}(t),\phi_{h}(t),\varrho_{h}(t)]=h[A(t),\phi(t),\varrho(t)]$.
Here $h$ is magnetic mirror transform. 

Assume the magnetic mirror transform for $\phi(t)$ and $\varrho(t)$
is same as to $E(t)$, The mirror transform of $A(t)$ is same as
$H$ or $B$. 

\[
[A_{h}(t),\phi_{h}(t),J_{h}(t),K_{h}(t),\varrho_{h}(t),t,\partial_{t}]\equiv h[A(t),\phi(t),J(t),K(t),\varrho(t)]
\]

\begin{equation}
=[-A(-t),\phi(-t),-J(-t),K(-t),\varrho(-t)]\label{eq:2100-80}
\end{equation}
It can be proven that the calculated $E_{h}(t)=-\nabla\phi_{h}-\partial A_{h}$,
$H_{h}(t)=\nabla\times A_{h}$ satisfies the magnetic mirrored transform
Eq.(\ref{eq:1000-20-1}). Here $h$ is magnetic mirror transform.
We need to prove that

1)
\begin{equation}
E_{h}(t)=E(-t)\label{eq:2100-90}
\end{equation}

2)
\begin{equation}
H_{h}(t)=-H(-t)\label{eq:2100-100}
\end{equation}

3)
\begin{equation}
\nabla\cdot J_{h}+\partial\varrho_{h}=0\label{eq:2100-110}
\end{equation}

Hence, guarantees the magnetic transformed field $[E_{h}(t),H_{h}(t)]$
satisfies the Maxwell equation Eq.(\ref{eq:490-1},\ref{eq:500-1}).
It also guarantees the current continue function still satisfies. 

Proof: After the magnetic mirror transform, the time $t$ is changed
to $-t$ and the $\partial_{t}$ change to $-\partial_{-t}=-\partial_{\tau}$,
here $-\tau=-t$ that is For example 

1)
\[
E_{h}(t)=-\nabla\phi_{h}-\partial_{t}A_{h}
\]
\[
=-\nabla\phi(-t)-\partial_{t}(-A(-t))
\]
\[
=-\nabla\phi(-t)-\partial_{-t}(A(-t))
\]
\[
=(-\nabla\phi(\tau)-\partial_{\tau}A(\tau))|_{\tau=-t}
\]
\[
=E(\tau)|_{\tau=-t}
\]
\begin{equation}
=E(-t)\label{eq:2100-145}
\end{equation}

2)
\begin{equation}
B_{h}=\nabla\times A_{h}=\nabla\times((-1)A(-t))=-\nabla\times A(-t)=-B(-t)\label{eq:2100-150}
\end{equation}
Hence
\begin{equation}
B_{h}=-B(-t)=-(\mu(\tau)*H(\tau))|_{\tau=-t}=-\mu(-t)*H(-t))=\mu_{h}*(-H(-t))\label{eq:2100-151}
\end{equation}
considering
\begin{equation}
B_{h}=\mu_{h}*H_{h}\label{eq:2100-152}
\end{equation}
Hence we have
\begin{equation}
H_{h}=-H(-t)\label{eq:2100-153}
\end{equation}

3)
\[
\nabla\cdot J_{h}+\partial\varrho_{h}
\]
\[
=\nabla\cdot(-J(-t))+\partial_{t}\varrho(-t))
\]
\[
=\nabla\cdot(-J(-t))-\partial_{-t}\varrho(-t))
\]
\[
=-(\nabla\cdot J(-t)+\partial_{-t}\varrho(-t))
\]
\[
=-(\nabla\cdot J(\tau)+\partial_{\tau}\varrho(\tau))
\]
\begin{equation}
=0\label{eq:210-160}
\end{equation}
In the above proof $\tau=-t$ has been used. Proof finish.

\subsection{Advanced potential and retarded potential}

In empty space the retarded potential is widely accept which are following,
\begin{equation}
E^{r}=-\nabla\phi^{r}-\partial A^{r}\label{eq:2100-10-1}
\end{equation}
\begin{equation}
B^{r}=\nabla\times A^{r}\label{eq:2100-20-1}
\end{equation}

\begin{equation}
\phi^{r}(x,t)=\frac{1}{4\pi\epsilon_{0}}\intop_{V}\frac{\varrho(x',t^{r})}{|x-x'|}dV\label{eq:2100-30}
\end{equation}
\begin{equation}
A^{r}(x,t)=\frac{\mu_{0}}{4\pi}\intop_{V}\frac{J(x',t^{r})}{|x-x'|}dV\label{eq:2100-40}
\end{equation}
\begin{equation}
t^{r}=t-\frac{|x-x'|}{c}\label{eq:2100-50}
\end{equation}
Where $c$ is the speed of light wave in empty space. Here $K=0$
is also assumed.

There is corresponding advanced potential, where

\begin{equation}
E^{a}=-\nabla\phi^{a}-\partial A^{a}\label{eq:2100-10-1-1}
\end{equation}
\begin{equation}
B^{a}=\nabla\times A^{a}\label{eq:2100-20-1-1}
\end{equation}

\begin{equation}
\phi^{a}(x,t)=\frac{1}{4\pi\epsilon_{0}}\intop_{V}\frac{\varrho(x',t^{a})}{|x-x'|}dV\label{eq:2100-30-1}
\end{equation}
\begin{equation}
A^{a}(x,t)=\frac{\mu_{0}}{4\pi}\intop_{V}\frac{J(x',t^{a})}{|x-x'|}dV\label{eq:2100-40-1}
\end{equation}

\begin{equation}
t^{a}=t+\frac{|x-x'|}{c}\label{eq:2100-50-1}
\end{equation}

The above electromagnetic field $\zeta^{r}=[E^{r},H^{r},J^{r},K^{r},\epsilon^{r},\mu^{r}]$
is the corresponding field of the retarded potential. $\zeta^{a}=[E^{a},H^{a},J^{a},K^{a},\epsilon^{a},\mu^{a}]$
is the corresponding field of the advanced potential. In the above
we can assume that $J^{r}=J^{a}=J$, $\epsilon^{r}=\epsilon^{a}=\epsilon$,
$\mu^{r}=\mu^{a}=\mu$. That means for the electric current $J$ and
media $\epsilon,\mu$, the superscript $r$ and $a$ can be dropped.

If the electric current or magnetic current $\rho=[J,K]$ sending
electromagnetic wave out, $\rho^{r}=[J^{r},K^{r}]$ is the source.
If the electric current or magnetic current $\rho^{a}=[J^{a},K^{a}]$
receiving electromagnetic wave, $\rho^{a}=[J^{a},K^{a}]$ is the sink.
The retarded potential is the electromagnetic field $\xi^{r}=[E^{r},H^{r}]$
transmitting from the source $\rho^{r}=[J^{r},K^{r}]$. The advanced
potential is the electromagnetic field $\xi=[E^{a},H^{a}]$ received
by the sink $\rho^{a}=[J^{a},K^{a}]$. Advanced potential and retarded
potential are all normal electromagnetic field which satisfies the
Maxwell equation. In next subsection we will shown that the field
of advanced potential is magnetic mirror transformed transformed field.
It is not the time reversed field. 

Here the mirror transformed field and time reversed field are different.
Mirror transformed fields still satisfy the Maxwell equation and hence
is a normal electromagnetic field. But the time reversed field does
not satisfy the Maxwell equation, it satisfies the time-reversed Maxwell
equation Eq.(\ref{eq:490-1-1-1-1}, \ref{eq:500-1-1-1-1}) which is
very close to Maxwell equation (It is noticed if the minus sign is
put to the media, the field after the time-reversed transform is still
satisfies the Maxwell equation). Time reversed transform can be apply
to derive some reciprocity theorems\cite{Samaddar}\cite{NorberN}.

\subsection{Obtain advanced potential from mirrored transform}

The retarded potential is widely accept. Advanced potential is not
widely accept. However magnetic mirror transformed field is accept
widely, since mirror transformed field satisfies the Maxwell equation
and also easy to explained as the reflect field on a magnetic super
conductor mirror. In this subsection we will derive the advanced potential
from magnetic mirror transform and retarded potential. 

Assume 
\begin{equation}
\zeta_{1}(t)=[E_{1}(t),H_{1}(t),J_{1}(t),K_{1}(t),\epsilon_{1}(t),\mu_{1}(t)]\label{eq:2110-10}
\end{equation}
\begin{equation}
\zeta_{2}(t)=[E_{2}(t),H_{2}(t),J_{2}(t),K_{2}(t),\epsilon_{2}(t),\mu_{2}(t)]\label{eq:2110-20}
\end{equation}
are both retarded potentials. Assume there is only electron current
hence $K_{1}=0$ and $K_{2}=0$. Assume 
\begin{equation}
[J_{2}(t),K_{2}(t),\epsilon_{2}(t),\mu_{2}(t),\varrho_{2}(t)]=[-J_{1}(-t),K_{1}(-t),\epsilon_{1}(-t),\mu_{1}(-t),\varrho_{1}(-t)]\label{eq:2110-25}
\end{equation}
corresponding to the source $\rho_{2}$ is the retarded potential,
and there is

\begin{equation}
\phi_{2}(x,t)=\frac{1}{4\pi\epsilon_{0}}\intop_{V}\frac{\varrho_{2}(x',t-\frac{|x-x'|}{c})}{|x-x'|}dV\label{eq:2110-30}
\end{equation}
\begin{equation}
A_{2}(x,t)=\frac{1}{4\pi\epsilon_{0}}\intop_{V}\frac{J_{2}(x',t-\frac{|x-x'|}{c})}{|x-x'|}dV\label{eq:2110-40}
\end{equation}
and the corresponding fields are
\begin{equation}
E_{2}=-\nabla\phi_{2}-\partial A_{2}\label{eq:2110-50}
\end{equation}
\begin{equation}
B_{2}=\nabla\times A_{2}\label{eq:2110-60}
\end{equation}
Assume 
\begin{equation}
\zeta_{3}=h\zeta_{2}\label{eq:2110-70}
\end{equation}
$\zeta_{3}$ is magnetic mirrored field of the retarded potential
$\zeta_{2}$. Considering 
\begin{equation}
A_{3}(x,t)=hA_{2}(x,t)=-A_{2}(x,-t)\label{eq:2110-80}
\end{equation}
\begin{equation}
\phi_{3}(x,t)=h\phi_{2}(x,t)=\phi_{2}(x,-t)\label{eq:2110-90}
\end{equation}
Considering Eq.(\ref{eq:2110-30},\ref{eq:2110-40}), there is

\begin{equation}
A_{3}(x,t)=(-)\frac{1}{4\pi\epsilon_{0}}\intop_{V}\frac{J_{2}(x',-t-\frac{|x-x'|}{c})}{|x-x'|}dV\label{eq:2110-100}
\end{equation}

\begin{equation}
\phi_{3}(x,t)=\frac{1}{4\pi\epsilon_{0}}\intop_{V}\frac{\varrho_{2}(x',-t-\frac{|x-x'|}{c})}{|x-x'|}dV\label{eq:2110-110}
\end{equation}
Considering Eq.(\ref{eq:2110-25}), there is

\begin{equation}
A_{3}(x,t)=\frac{1}{4\pi\epsilon_{0}}\intop_{V}\frac{J_{1}(x',t+\frac{|x-x'|}{c})}{|x-x'|}dV\label{eq:2110-120}
\end{equation}
\begin{equation}
\phi_{3}(x,t)=\frac{1}{4\pi\epsilon_{0}}\intop_{V}\frac{\varrho_{1}(x',t+\frac{|x-x'|}{c})}{|x-x'|}dV\label{eq:2110-130}
\end{equation}
The field can be obtained

\begin{equation}
E_{3}=-\nabla\phi_{3}-\partial A_{3}\label{eq:2100-140}
\end{equation}
\begin{equation}
B_{3}=\nabla\times A_{3}\label{eq:2100-200}
\end{equation}
This way It is proven that the magnetic mirror transformed field 
\begin{equation}
\zeta_{3}(t)=[E_{3}(t),H_{3}(t),J_{3}(t),K_{3}(t),\epsilon_{3}(t),\mu_{3}(t)]\label{eq:2100-2011}
\end{equation}
 from a field of the retarded potential 
\begin{equation}
\zeta_{2}(t)=[E_{2}(t),H_{2}(t),J_{2}(t),K_{2}(t),\epsilon_{2}(t),\mu_{2}(t)]\label{eq:2100-2012}
\end{equation}
 is just the field of the advanced potential of the sink $\rho_{1}=[J_{1}(t),0]$. 

The field of the advanced potential is obtained by using a magnetic
mirror transform. Hence the advanced potential should be acceptable
same as the field obtained form a magnetic mirror transform. In the
following the transmitting field and retarded potential are field
that send out from the source $\rho=[J,K]$. Receiving potential and
advanced potential are field receiving by the sink $\rho=[J,K]$.
Electric and magnetic current can receive and transmit the field or
do both in the same time, i.e. $\rho=[J,K]$ can be sink or source.

\section{The modified Poynting theorem}

\subsection{The superimposing electromagnetic field}

The superimposing electromagnetic field is considered which contains
the following electromagnetic field compounds,

(1) retarded potential which send from the source.

(2) advanced potential which is received by the sink.

(3) time reversed field of the retarded potential and advanced potential.

(4) mirror transformed field of the retarded potential and advanced
potential.

(5) time-offset field of the above fields.

(6) space-offset field of the above fields.

(7) transmitting field, which is sent from the antenna.

(8) receiving field, which receiving from antenna and reflected by
the antenna

The Maxwell equation Eq.(\ref{eq:490-1}, \ref{eq:500-1}) satisfies.
This is also referred as the modified Poynthing theorem. The simple
example is shown in the following. Assume $\zeta_{1}$, $\zeta_{2}$,$\zeta_{3}$,$\zeta_{4},\zeta_{5}$,$\zeta_{6}$
are all the field of the retarded potential, 
\begin{equation}
\zeta=\zeta_{1}+h\zeta_{2}+e\zeta_{3}+T\zeta_{4}+X\zeta_{5}+r\zeta_{6}\label{eq:1000-40}
\end{equation}
is superimposed field. Where $h$ and $e$ are magnetic and electric
mirror transforms. $h\zeta_{2}$ and $e\zeta_{3}$ are receiving fields
or advanced potential. $T$ is time offset transform, $X$ is space
offset transform. $T\zeta_{4}$ is time-offset field, $X\zeta_{5}$
is spatial offset fields. $r\zeta_{6}$ is time reversed transform.

\subsection{The Poynting theorem}

The Poynting theorem can be proved as following from Maxwell equation
Eq.(\ref{eq:490-1},\ref{eq:500-1}),

\begin{equation}
-\nabla\cdot(E\times H)=J\cdot E+K\cdot H+E\cdot\partial D+H\cdot\partial B\label{eq:30}
\end{equation}
Where ``$\cdot$'' is vector point product. The superimposed field
Eq.(\ref{eq:1000-40}) satisfies the Maxwell equation, if the media
satisfies Eq.(\ref{eq:501-1-2-1-3},\ref{eq:502-1-2-1-3}),i.e., there
is only one media, $[\epsilon,\mu]$ we say that the Poynting theorem
establishes.

\subsection{The modified Poynting theorem}

The Eq.(\ref{eq:30}) is also the modified Poynting theorem if we
consider the media Eq.(\ref{eq:501-1-2-1-1},\ref{eq:502-1-2-1-1}).
The derivation of the modified Poynting theorem can be done with Maxwell
equation and the modified media Eq.(\ref{eq:501-1-2-1-1},\ref{eq:502-1-2-1-1}).
The derivation of the modified Poynting theorem is exactly same as
derivation Poynting theorem form Maxwell equation. The only different
is that the media have been generalized to Eq.(\ref{eq:501-1-2-1-1},\ref{eq:502-1-2-1-1})
from Eq.(\ref{eq:501-1-2-1-3},\ref{eq:502-1-2-1-3}). If Eq.(\ref{eq:501-1-2-1-3},\ref{eq:502-1-2-1-3})
is applied, the word ``modified'' can be dropped, it become Poynting
theorem. The concept of ``modified'' is borrowed from the modified
reciprocity theorem\cite{J_A_Kong2}. We extended this idea to the
Maxwell equation and also Poynting theorem.

\subsection{The Poynting theorem in Fourier space}

Considering 
\[
f(t)=F^{-1}(f(\omega))=\frac{1}{2\pi}\intop_{\omega=-\infty}^{+\infty}f(\omega)\exp(j\omega t)\, d\omega
\]
\[
\partial f(t)=\frac{1}{2\pi}\intop_{\omega=-\infty}^{+\infty}f(\omega)\partial\exp(j\omega t)\, d\omega
\]
\[
=\frac{1}{2\pi}\intop_{\omega=-\infty}^{+\infty}f(\omega)(j\omega)\exp(j\omega t)\, d\omega
\]
Hence there is
\[
F\{\partial f(t)\}=(j\omega)f(\omega)
\]
or there is the transform

\[
\partial\rightarrow(j\omega)
\]
from time-domain to Frequency domain. And the Maxwell equation in
Fourier space becomes,

\begin{equation}
\nabla\times H=J+(j\omega)D\label{eq:490-1-3}
\end{equation}
\begin{equation}
\nabla\times E=-K-(j\omega)B\label{eq:500-1-3}
\end{equation}
In frequency domain the Poynting theorem Eq.(\ref{eq:30}) can be
written as

\begin{equation}
-\nabla\cdot(E\times H)=J\cdot E+K\cdot H+E\cdot j\omega D+H\cdot j\omega B\label{eq:30-1}
\end{equation}

\subsection{The complex Poynting theorem}

Considering $\partial_{t}\rightarrow j\omega$ in the Maxwell equation
Eq.(\ref{eq:490-1},\ref{eq:500-1}), there is the Maxwell equation
in the Fourier domain,

\begin{equation}
\nabla\times H=J+(j\omega)\epsilon E\label{eq:9000-10}
\end{equation}

\begin{equation}
-\nabla\times E=K+j\omega\mu H\label{eq:9000-20}
\end{equation}
take complex conjugate to Eq.(\ref{eq:9000-10}), there is

\begin{equation}
\nabla\times H^{*}=J^{*}+(j\omega)^{*}\epsilon^{*}E^{*}\label{eq:9000-30}
\end{equation}
point product a variable $H^{*}$ and $E$ to Eq.(\ref{eq:9000-20},\ref{eq:9000-30}),
there are
\begin{equation}
-H^{*}\cdot\nabla\times E=H^{*}\cdot K+j\omega H^{*}\cdot\mu H\label{eq:9000-40}
\end{equation}
\begin{equation}
E\cdot\nabla\times H^{*}=E\cdot J^{*}+(j\omega)^{*}E\cdot\epsilon^{*}E^{*}\label{eq:9000-50}
\end{equation}
Add them together
\begin{equation}
-H^{*}\cdot\nabla\times E+E\cdot\nabla\times H^{*}=H^{*}\cdot K+j\omega H^{*}\cdot\mu H+E\cdot J^{*}+(j\omega)^{*}E\cdot\epsilon^{*}E^{*}\label{eq:9000-60}
\end{equation}
Poynting theorem in Fourier domain can be written as

\begin{equation}
-\nabla\cdot(E\times H^{*})=E\cdot J^{*}+H^{*}\cdot K+j\omega(H^{*}\cdot\mu H-E\cdot\epsilon^{*}E^{*})\label{eq:9000-70}
\end{equation}

\section{Fail to derive the mutual energy theorem from complex Poynting theorem}

Our first try is to obtained mutual energy theorem from complex Poynting
theorem. This try is failed, instead of obtained the mutual energy
theorem we obtained the mixed mutual energy theorem. 

Assume $\zeta=\zeta_{1}+\zeta_{2}$, the above formula can be rewritten
as
\[
-\nabla\cdot((E_{1}+E_{2})\times(H_{1}^{*}+H_{2}^{*}))
\]
\[
=(E_{1}+E_{2})\cdot(J_{1}^{*}+J_{2}^{*})+(H_{1}^{*}+H_{2}^{*})\cdot(K_{1}+K_{2})
\]
\begin{equation}
+j\omega((H_{1}^{*}+H_{2}^{*})\cdot(\mu_{1}H_{1}+\mu_{2}H_{2})-(E_{1}+E_{2})\cdot(\epsilon_{1}^{*}E_{1}^{*}+\epsilon_{2}^{*}E_{2}^{*}))\label{eq:9000-80}
\end{equation}
take out all self energy compounds, it becomes

\[
-\nabla\cdot(E_{1}\times H_{2}^{*}+E_{2}\times H_{1}^{*})
\]
\[
=E_{1}\cdot J_{2}^{*}+E_{2}\cdot J_{1}^{*}+H_{1}^{*}\cdot K_{2}+H_{2}^{*}\cdot K_{1}
\]
\begin{equation}
+j\omega(H_{1}^{*}\cdot\mu_{2}H_{2}+H_{2}^{*}\cdot\mu_{1}H_{1}-E_{1}\cdot\epsilon_{2}^{*}E_{2}^{*}-E_{2}\cdot\epsilon_{1}^{*}E_{1}^{*})\label{eq:9000-90}
\end{equation}
or in intergral form,

\[
-\intop_{S}(E_{1}\times H_{2}^{*}+E_{2}\times H_{1}^{*})\cdot\hat{n}dS
\]
\[
=\intop_{V}(E_{1}\cdot J_{2}^{*}+E_{2}\cdot J_{1}^{*}+H_{1}^{*}\cdot K_{2}+H_{2}^{*}\cdot K_{1})dV
\]
\begin{equation}
+j\omega\intop_{V}(H_{1}^{*}\cdot\mu_{2}H_{2}+H_{2}^{*}\cdot\mu_{1}H_{1}-E_{1}\cdot\epsilon_{2}^{*}E_{2}^{*}-E_{2}\cdot\epsilon_{1}^{*}E_{1}^{*})dV\label{eq:9000-100}
\end{equation}
This is referred as mixed mutual energy theorem. It is related to
the concept of mixed Poynting vector. It is corresponding to Poynting
theorem in the complex form. The inverse Fourier transform of above
formula is
\[
-\intop_{S}\intop_{t=-\infty}^{\infty}(E_{1}(t+\tau)\times H_{2}^{*}(t)+E_{2}(t+\tau)\times H_{1}^{*}(\tau))dt\cdot\hat{n}dS
\]
\[
=\intop_{V}\intop_{t=-\infty}^{\infty}(E_{1}(t+\tau)\cdot J_{2}^{*}(t)+E_{2}(t+\tau)\cdot J_{1}^{*}(t)+H_{1}^{*}(t)\cdot K_{2}(t+\tau)+H_{2}^{*}(t)\cdot K_{1}(t+\tau))\, dt\, dV
\]
\[
+\partial_{\tau}\intop_{V}\intop_{t=-\infty}^{\infty}(H_{1}^{*}(t)\cdot(\mu_{2}*H_{2})(t+\tau)+H_{2}^{*}(t)\cdot(\mu_{1}*H_{1})(t+\tau)
\]

\begin{equation}
-E_{1}(t+\tau)\cdot(\epsilon_{2}^{*}*E_{2}^{*})(t)-E_{2}(t+\tau)\cdot(\epsilon_{1}^{*}*E_{1}^{*})(t))\, dt\, dV\label{eq:9000115}
\end{equation}
In the above formula ($f*g)(t)$ means convolution of the function
$f(t)$ and $g(t)$. Considering $E(t),H(t),\epsilon(t),\mu(t)$ are
all real variables, the above formula can be rewritten as following,
\[
-\intop_{S}(E_{1}(t+\tau)\times H_{2}(t)+E_{2}(t+\tau)\times H_{1}(\tau))\cdot\hat{n}dS
\]
\[
=\intop_{V}(E_{1}(t+\tau)\cdot J_{2}(t)+E_{2}(t+\tau)\cdot J_{1}(t)+H_{1}(t)\cdot K_{2}(t+\tau)+H_{2}(t)\cdot K_{1}(t+\tau))dV
\]
\[
+\partial_{\tau}\intop_{V}(H_{1}(t)\cdot(\mu_{2}*H_{2})(t+\tau)+H_{2}(t)\cdot(\mu_{1}*H_{1})(t+\tau)
\]
\begin{equation}
-E_{1}(t+\tau)\cdot(\epsilon_{2}*E_{2})(t)-E_{2}(t+\tau)\cdot(\epsilon_{1}*E_{1})(t))dV\label{eq:9000-120}
\end{equation}
This can be seen as mixed mutual energy theorem in time domain, it
also can be referred as mixed time-correlation reciprocity theorem.
It can be prove that the the real part of mixed mutual energy theorem
Eq.(\ref{eq:9000-100}) is same as the real part of the mutual energy
theorem Eq.(\ref{eq:2000-140}). Mixed mutual energy theorem is related
to the concept of mixed poynting vector\cite{Fragstein-Conrad,Angus-Macleod}.

\section{Derivation of mutual energy theorem by average}

We define the mutual energy of a electromagnetic field system as the
difference between the total energy and the self energy. We derive
mutual energy from the idea that subtract the self energy from the
total energy, the rest is the mutual energy. There is energy conservation
theorem which is Poynting theorem which guarantee the total energy
and self energy are conservation. Hence the mutual energy should also
be conserved. The mutual energy is conserved is referred as mutual
energy theorem. The following gives the detail of the derivation of
the mutual energy theorem

\subsection{Spatial-temporal mutual energy theorem}

Assume 
\begin{equation}
E=E_{1}+E_{2}\label{eq:40}
\end{equation}
\begin{equation}
D=D_{1}+D_{2}\label{eq:40-1}
\end{equation}

\begin{equation}
H=H_{1}+H_{2}\label{eq:50}
\end{equation}
\begin{equation}
B=B_{1}+B_{2}\label{eq:50-1}
\end{equation}

\begin{equation}
J=J_{1}+J_{2}\label{eq:60}
\end{equation}
Assume $(E_{1},H_{1})$ is produced from $J_{1}$ and assume $(E_{2},H_{2})$
is produced from $J_{2}$. It is clear we have the Poynting theorem
as following, 
\begin{equation}
-\nabla\cdot(E_{1}\times H_{1})=J_{1}\cdot E_{1}+E_{1}\cdot\partial D_{1}+H_{1}\cdot\partial B_{2}\label{eq:70-2}
\end{equation}

\begin{equation}
-\nabla\cdot(E_{2}\times H_{2})=J_{2}\cdot E_{2}+E_{2}\cdot\partial D_{2}+H\cdot\partial B_{2}\label{eq:80-1}
\end{equation}
substitute Eq.(\ref{eq:40},\ref{eq:40-1},\ref{eq:50},\ref{eq:50-1})
to Eq.(\ref{eq:30}) and subtract Eq.(\ref{eq:70-2}) and Eq.(\ref{eq:80-1})
from Eq.(\ref{eq:30}) , we obtain,
\begin{equation}
-\nabla\cdot(E_{1}\times H_{2}+E_{2}\times H_{1})=J_{1}\cdot E_{2}+J_{2}\cdot E_{1}+E_{1}\cdot\partial D_{2}+E_{2}\cdot\partial D_{1}+H_{1}\cdot\partial B_{2}+H_{2}\cdot\partial B_{1}\label{eq:90-2}
\end{equation}
Actually we can call Eq.(\ref{eq:30}) total part of Poynting theorem.
The Eq.(\ref{eq:70-2},\ref{eq:80-1}) is self part of Poynting theorem.
Eq.(\ref{eq:90-2}) is mutual part of Poynting theorem. The corresponding
integral form of the above formula is 

\begin{equation}
-\intop\intop_{S}(E_{1}\times H_{2}+E_{2}\times H_{1})\cdot\hat{n}dS=\intop\intop\intop_{V}\,(J_{1}\cdot E_{2}+J_{2}\cdot E_{1}+E_{1}\cdot\partial D_{2}+E_{2}\cdot\partial D_{1}+H_{1}\cdot\partial B_{2}+H_{2}\cdot\partial B_{1})\, dV\label{eq:100-1}
\end{equation}
Where $\hat{n}$ is norm vector of the the surface $S$. The above
Eq.(\ref{eq:90-2},\ref{eq:100-1}) are referred as spatial-temporal
mutual energy theorem in this article. In the derivation of the above
Eq.(\ref{eq:90-2},\ref{eq:100-1}) we have considered the media,

\begin{equation}
D_{1}=\epsilon E_{1}\label{eq:110}
\end{equation}
\begin{equation}
D_{2}=\epsilon E_{2}\label{eq:120}
\end{equation}
\begin{equation}
B_{1}=\mu H_{1}\label{eq:130}
\end{equation}
\begin{equation}
B_{2}=\mu H_{2}\label{eq:140}
\end{equation}

\subsection{Modified spatial-temporal mutual energy theorem}

If we assume that

\begin{equation}
D_{1}=\epsilon_{1}E_{1}\label{eq:110-1}
\end{equation}
\begin{equation}
D_{2}=\epsilon_{2}E_{2}\label{eq:120-1}
\end{equation}
\begin{equation}
B_{1}=\mu_{2}H_{1}\label{eq:130-1}
\end{equation}
\begin{equation}
B_{2}=\mu_{2}H_{2}\label{eq:140-1}
\end{equation}
Eq.(\ref{eq:90-2},\ref{eq:100-1}) are referred as modified spatial-temporal
mutual energy theorem. Modified spatial-temporal mutual energy theorem
can be derived directly from modified Poynting theorem. We can see
in the derivation if only consider $E$, $H$, $D$, $B$, $J$. The
medium $\epsilon_{1}$,$\mu_{1}$ and $\epsilon_{2}$,$\mu_{2}$ did
not appear in the proving process. Hence using the exactly same proving
process we can obtain the modified spatial-temporal mutual energy
theorem. 

It is remarkable that we did not care whether or not the modified
spatial-temporal mutual energy theorem is a true physical theorem,
we can consider it as a mathematical theorem. In real situation there
is only 
\begin{equation}
\epsilon_{1}=\epsilon_{2}=\epsilon\label{eq:150}
\end{equation}
\begin{equation}
\mu_{1}=\mu_{2}=\mu\label{eq:160}
\end{equation}
Modified mutual energy theorem can be used to simplify some calculation
of electromagnetic fields which will be shown in the following.

\subsection{Mutual energy theorem in complex space}

Considering $f=Re\{f_{0}\exp(j\omega t)\}$ $g=Re\{g_{0}\exp(j\omega t)\}$,
\[
f\, g=Re\{f_{0}\exp(j\omega t)\}Re\{g_{0}\exp(j\omega t)\}
\]
\[
=\frac{1}{2}(f_{0}\exp(j\omega t)+f_{0}^{*}(\exp(j\omega t))^{*})\frac{1}{2}(g_{0}\exp(j\omega t)+g_{0}^{*}(\exp(j\omega t))^{*}
\]
\[
=\frac{1}{4}(f_{0}g_{0}\exp(j2\omega t)+f_{0}^{*}g_{0}^{*}(\exp(j2\omega t))^{*}+f_{0}g_{0}^{*}+f_{0}^{*}g_{0})
\]
\begin{equation}
=\frac{1}{2}(Re\{f_{0}g_{0}\exp(j2\omega t)\}+Re\{f_{0}g_{0}^{*}\})\label{eq:161}
\end{equation}
Considering
\begin{equation}
<f_{0}g_{0}\exp(j2\omega t)>=0\label{eq:162}
\end{equation}
Where$<\bullet>$means average of some variable with time. Hence we
have

\[
<f,g>=\frac{1}{2}Re\{f_{0}g_{0}^{*}\}
\]
\[
=\frac{1}{2}Re\{f_{0}\exp(j\omega t)\, g_{0}^{*}(\exp(j\omega t))^{*}\}
\]
\begin{equation}
=\frac{1}{2}Re\{f\, g^{*})\}\label{eq:163}
\end{equation}
Hence we have the following transform after the average if do not
consider the constant $\frac{1}{2}$, 
\begin{equation}
f\, g\Longrightarrow f\, g^{*}\label{eq:164}
\end{equation}
or $\Longrightarrow$means take an average

\begin{equation}
f\, g\Longrightarrow f^{*}\, g\label{eq:168}
\end{equation}
According to this, we have 

\begin{equation}
E_{1}\times H_{2}\Longrightarrow E_{1}\times H_{2}^{*}\label{eq:250}
\end{equation}

\begin{equation}
E_{2}\times H_{1}\Longrightarrow E_{2}^{*}\times H_{1}\label{eq:260}
\end{equation}

\begin{equation}
J_{1}\times E_{2}\Longrightarrow J_{1}\times E_{2}^{*}\label{eq:270}
\end{equation}
\begin{equation}
J_{2}\times E_{1}\Longrightarrow J_{2}^{*}\times E_{1}\label{eq:280}
\end{equation}
In the above substitution we always put the $*$ to the all variable
with subscript $2$. considering,

\[
\partial f=\partial Re\{f_{0}\exp(j\omega t)\}
\]
\[
=\partial\frac{1}{2}(f_{0}\exp(j\omega t)+f_{0}^{*}(\exp(j\omega t))^{*})
\]
\[
=\frac{1}{2}(j\omega f_{0}\exp(j\omega t)+f_{0}^{*}(-j\omega)(\exp(j\omega t))^{*})
\]
\[
=\frac{1}{2}Re\{(j\omega f_{0}\exp(j\omega t)\}
\]
\begin{equation}
=\frac{1}{2}Re\{(j\omega f\}\label{eq:281}
\end{equation}
Hence we have the following transform

\begin{equation}
\partial\Longrightarrow j\omega\label{eq:282}
\end{equation}
We obtain,
\begin{equation}
E_{1}\cdot\partial D_{2}\Longrightarrow E_{1}\cdot(\partial\epsilon E_{2})^{*}=E_{1}\cdot(j\omega\epsilon E_{2})^{*}=(j\omega)^{*}\, E_{1}\cdot\epsilon^{*}E_{2}^{*}=-j\omega\, E_{1}\cdot\epsilon^{*}E_{2}^{*}\label{eq:290}
\end{equation}
and
\begin{equation}
E_{2}\cdot\partial D_{1}\Longrightarrow E_{2}^{*}\cdot(j\omega\epsilon E_{1})=j\omega E_{2}^{*}\cdot\epsilon E_{1}\label{eq:300}
\end{equation}
similarly we have

\begin{equation}
H_{1}\cdot\partial B_{2}\Longrightarrow-j\omega\, H_{1}\cdot\mu^{*}H_{2}^{*}\label{eq:310}
\end{equation}
\begin{equation}
H_{2}\cdot\partial B_{1}\Longrightarrow j\omega H_{2}^{*}\cdot\mu B_{1}\label{eq:320}
\end{equation}
 Substitute the above formula to Eq.(\ref{eq:90-2}) we have 
\[
-\nabla\cdot(E_{1}\times H_{2}^{*}+E_{2}^{*}\times H_{1})=
\]
\begin{equation}
J_{1}\cdot E_{2}^{*}+J_{2}^{*}\cdot E_{1}-j\omega\, E_{1}\cdot\epsilon^{*}E_{2}^{*}+j\omega E_{2}^{*}\cdot\epsilon E_{1}-j\omega\, H_{1}\cdot\mu^{*}H_{2}^{*}+j\omega H_{2}^{*}\cdot\mu B_{1}\label{eq:330}
\end{equation}
Considering
\begin{equation}
a\cdot Cb=a_{i}C_{ij}b_{j}=C_{ij}a_{i}b_{j}=C_{ji}^{T}a_{i}\cdot b_{j}=b\cdot C^{T}a\label{eq:350}
\end{equation}
Hence we have
\begin{equation}
E_{1}\cdot\epsilon^{*}E_{2}^{*}=E_{2}^{*}\cdot(\epsilon^{*})^{T}E_{1}=E_{2}^{*}\cdot\epsilon^{\dagger}E_{1}\label{eq:360}
\end{equation}
Where $A^{\dagger}=(A^{*})^{T}$. $\dagger$ means matrix conjugate.
$T$ means matrix transpose. Hence we have
\begin{equation}
E_{1}\epsilon^{*}E_{2}^{*}-E_{2}^{*}\epsilon E_{1}=E_{2}^{*}\epsilon^{\dagger}E_{1}-E_{2}^{*}\epsilon E_{1}=E_{2}^{*}(\epsilon^{\dagger}-\epsilon)E_{1}\label{eq:370}
\end{equation}
\begin{equation}
H_{1}\mu^{*}H_{2}^{*}-H_{2}^{*}\mu H_{1}=H_{2}^{*}(\mu^{\dagger}-\mu)H_{1}\label{eq:370-1}
\end{equation}
Hence Eq.(\ref{eq:330}) can be rewritten as following,
\begin{equation}
-\nabla\cdot(E_{1}\times H_{2}^{*}+E_{2}^{*}\times H_{1})=J_{1}\cdot E_{2}^{*}+J_{2}^{*}\cdot E_{1}-j\omega E_{2}^{*}\cdot(\epsilon^{\dagger}-\epsilon)E_{1}-j\omega H_{2}^{*}\cdot(\mu^{\dagger}-\mu)H_{1}\label{eq:380}
\end{equation}
The corresponding integral form is
\begin{equation}
-\intop\intop_{S}(E_{1}\times H_{2}^{*}+E_{2}^{*}\times H_{1})\cdot\hat{n}dS=\intop\intop\intop_{V}(J_{1}\cdot E_{2}^{*}+J_{2}^{*}\cdot E_{1}-j\omega E_{2}^{*}\cdot(\epsilon^{\dagger}-\epsilon)E_{1}-j\omega H_{2}^{*}\cdot(\mu^{\dagger}-\mu)H_{1})\, dv\label{eq:390}
\end{equation}

\subsection{mutual energy theorem in complex space with lossless medium}

If the medium is lossless, there are following condition, 
\begin{equation}
\epsilon^{\dagger}=\epsilon\label{eq:420}
\end{equation}
\begin{equation}
\mu^{\dagger}=\mu\label{eq:430}
\end{equation}
Hence there is

\begin{equation}
j\omega E_{2}^{*}\cdot(\epsilon^{\dagger}-\epsilon)E_{1}+j\omega H_{2}^{*}\cdot(\mu^{\dagger}-\mu)H_{1}=0\label{eq:431}
\end{equation}
Hence the complex mutual energy theorem can be simplified to the following
form,

\begin{equation}
-\nabla\cdot(E_{1}\times H_{2}^{*}+E_{2}^{*}\times H_{1})=J_{1}\cdot E_{2}^{*}+J_{2}^{*}\cdot E_{1}\label{eq:440}
\end{equation}
or if the integral form,
\begin{equation}
-\intop\intop_{S}(E_{1}\times H_{2}^{*}+E_{2}^{*}\times H_{1})\cdot\hat{n}dS=\intop\intop\intop_{V}(J_{1}\cdot E_{2}^{*}+J_{2}^{*}\cdot E_{1})\, dV\label{eq:450}
\end{equation}
This is referred as simplified form of the mutual energy theorem.
Or for simplification, just mutual energy theorem\cite{shrzhao1,shrzhao2,shrzhao3}.
It has been referred as the second reciprocity theorem\cite{Petrusenko},
generalized reciprocity theorem, adjoint reciprocity theorem, lossless
reciprocity theorem.

\subsection{Modified mutual energy theorem in complex space}

The above are referred as the complex mutual energy theorem. The corresponding
modified complex mutual energy theorem is 

\begin{equation}
-\nabla\cdot(E_{1}\times H_{2}^{*}+E_{2}^{*}\times H_{1})\cdot\hat{n}dS=J_{1}\cdot E_{2}^{*}+J_{2}^{*}\cdot E_{1}-j\omega E_{2}^{*}\cdot(\epsilon_{2}^{\dagger}-\epsilon_{1})E_{1}-j\omega H_{2}^{*}\cdot(\mu_{2}^{\dagger}-\mu_{1})H_{1}\label{eq:400}
\end{equation}
\begin{equation}
-\intop\intop_{S}(E_{1}\times H_{2}^{*}+E_{2}^{*}\times H_{1})\cdot\hat{n}dS=\intop\intop\intop_{V}(J_{1}\cdot E_{2}^{*}+J_{2}^{*}\cdot E_{1}-j\omega E_{2}^{*}\cdot(\epsilon_{2}^{\dagger}-\epsilon_{1})E_{1}-j\omega H_{2}^{*}\cdot(\mu_{2}^{\dagger}-\mu_{1})H_{1})\, dV\label{eq:410}
\end{equation}

\subsection{Modified mutual energy theorem in complex space with loss medium}

If the lossless condition does not satisfy. The mutual energy formula
can not be written as the above simplified form. However we always
can choose 

\begin{equation}
\epsilon_{2}^{\dagger}=\epsilon_{1}\label{eq:460}
\end{equation}
\begin{equation}
\mu_{2}^{\dagger}=\mu_{1}\label{eq:470}
\end{equation}
So that the Eq.(\ref{eq:440},\ref{eq:450}) are still true in the
meaning of modified mutual energy theorem. We must keep in mind that
Eq.(\ref{eq:440},\ref{eq:450}) are both the mutual energy theorem
and the modified mutual energy theorem depending whether or not the
medium i.e. the Eq.(\ref{eq:150},\ref{eq:160}) satisfy or not. Here
usually one of media is in the real space for example $\epsilon_{1}$,
$\mu_{1}$, the other $\epsilon_{2}$ are $\mu_{2}$ are in virtual
space and it is effective only with mathematical meaning. In this
situation we can free to choose their values. The simplified form
of the modified mutual energy theorem can be found in reference\cite{shrzhao1,shrzhao2,shrzhao3}
in which is just called as modified mutual energy theorem.

It is worth to see that the above (modified) mutual energy theorem
is derived from modified Poynting theorem, time-offset, time-reversed,
space-offset.

\section{Derive the theorems in Fourier domain directly\label{sec:Derive-the-theorems}}

In the last section we have derived the mutual energy theorem from
complex Poynting theorem using the process of average. However that
derivation is not strictly. Since actually we have only proved the
real part of the theorem. The mutual energy theorem has image part.
We have not prove the image part of mutual energy theorem. In this
section we will solve this problem.

\subsection{The modified reversed mutual energy theorem}

Assume $\zeta_{1}=[E_{1},H_{1},J_{1},K_{1},D_{1},B_{1}]$, $\zeta_{2}=[E_{2},H_{2},J_{2},K_{2},D_{2},B_{2}]$,
$\zeta=\zeta_{1}+\zeta_{2}$, there is the total energy formula, 
\[
-\nabla\cdot((E_{1}+E_{2})\times(H_{1}+H_{2}))
\]

\begin{equation}
=(J_{1}+J_{2})\cdot(E_{1}+E_{2})+(K_{1}+K_{2})\cdot(H_{1}+H_{2})+(E_{1}+E_{2})\cdot j\omega(D_{1}+D_{2})+(H_{1}+H_{2})\cdot j\omega(B_{1}+B_{2})\label{eq:30-1-1}
\end{equation}
And the self energy formula,
\begin{equation}
-\nabla\cdot(E_{1}\times H_{1})=J_{1}\cdot E_{1}+K_{1}\cdot H_{1}+E_{1}\cdot j\omega D_{1}+H_{1}\cdot j\omega B_{1}\label{eq:30-1-2}
\end{equation}

\begin{equation}
-\nabla\cdot(E_{2}\times H_{2})=J_{2}\cdot E_{2}+K_{2}\cdot H_{2}+E_{2}\cdot j\omega D_{2}+H_{2}\cdot j\omega B_{2}\label{eq:30-1-3}
\end{equation}
Subtract the above two self energy formulas form the total energy
formula, we obtain

\[
-\nabla\cdot(E_{1}\times H_{2}+E_{2}\times H_{1})
\]

\begin{equation}
=J_{1}\cdot E_{2}+J_{2}\cdot E_{1}+K_{1}\cdot H_{2}+K_{2}\cdot H_{1}+j\omega(E_{1}\cdot D_{2}+E_{2}\cdot D_{1}+H_{1}\cdot B_{2}+H_{2}\cdot B_{1})\label{eq:30-1-1-1}
\end{equation}
Among the above formula we notice that
\[
E_{1}\cdot D_{2}+E_{2}\cdot D_{1}=E_{1}\cdot\epsilon_{2}E_{2}+E_{2}\cdot\epsilon_{1}E_{1}
\]
\[
=E_{1}\cdot\epsilon_{2}E_{2}+\cdot E_{1}\epsilon_{1}^{T}E_{2}
\]
\begin{equation}
=E_{1}\cdot(\epsilon_{2}+\epsilon_{1}^{T})E_{2}\label{eq:7200-200}
\end{equation}
and

\begin{equation}
H_{1}\cdot B_{2}+H_{2}\cdot B_{1}=H_{1}\cdot(\mu_{2}+\mu_{1}^{T})H_{2}\label{eq:7200-210}
\end{equation}
Hence if we choose $\epsilon_{2}(\omega)$ and $\mu_{2}(\omega)$
satisfy 
\begin{equation}
\epsilon_{2}(\omega)+\epsilon_{1}^{T}(\omega)=0,\ \ \ \ \ \ \ \ \ \mu_{2}(\omega)+\ \mu_{1}^{T}(\omega)=0\label{eq:7200-230}
\end{equation}
There is
\[
-\nabla\cdot(E_{1}(\omega)\times H_{2}(\omega)+E_{2}(\omega)\times H_{1}(\omega))
\]

\begin{equation}
=J_{1}(\omega)\cdot E_{2}(\omega)+E_{1}(\omega)\cdot J_{2}(\omega)+K_{1}(\omega)\cdot H_{2}(\omega)+H_{1}(\omega)\cdot K_{2}(\omega)\label{eq:7200-240}
\end{equation}
the integral form form can be written as 
\[
-\intop_{S}(E_{1}(\omega)\times H_{2}(\omega)+E_{2}(\omega)\times H_{1}(\omega))\:\hat{n}dS
\]
\begin{equation}
=\intop_{V}(J_{1}(\omega)\cdot E_{2}(\omega)+E_{1}(\omega)\cdot J_{2}(\omega)+K_{1}(\omega)\cdot H_{2}(\omega)+H_{1}(\omega)\cdot K_{2}(\omega))\, dV\label{eq:7200-250}
\end{equation}
The above two formula can be referred as the modified reversed mutual
theorem. Define

\begin{equation}
(\xi_{1},\xi_{2})_{r\omega}=\intop_{S}(E_{1}(\omega)\times H_{2}(\omega)+E_{2}(\omega)\times H_{1}(\omega))\:\hat{n}dS\label{eq:7200-260}
\end{equation}
\begin{equation}
(\rho_{1},\xi_{2})_{r\omega}=\intop_{V}(J_{1}(\omega)\cdot E_{2}(\omega)+K_{1}(\omega)\cdot H_{2}(\omega))dV\label{eq:7200-270}
\end{equation}
\begin{equation}
(\xi_{1},\rho_{2})_{r\omega}=\intop_{V}(E_{1}(\omega)\cdot J_{2}(\omega)+H_{1}(\omega)\cdot K_{2}(\omega))dV\label{eq:7200-280}
\end{equation}
Subscript ``$r$'' is used to express this inner product has not
use any complex conjugate symbols. We have the reversed mutual energy
theorem as following,
\begin{equation}
(\xi_{1},\xi_{2})_{r\omega}+(\rho_{1},\xi_{2})_{r\omega}+(\xi_{1},\rho_{2})_{r\omega}=0\label{eq:7200-290}
\end{equation}

\begin{equation}
\epsilon_{2}(\omega)+\epsilon_{1}^{T}(\omega)=0,\ \ \ \ \ \ \ \ \ \mu_{2}(\omega)+\mu_{1}^{T}(\omega)=0\label{eq:7200-300}
\end{equation}
In the history, the closed work related Eq.(\ref{eq:7200-250}) can
be found in reference.\cite{Samaddar} or see Eq.(\ref{eq:10000-50}).
In case $\epsilon_{1}=\epsilon_{2}=\epsilon$, $\mu_{1}=\mu_{2}=\mu$,
the word ``modified'' can be dropped. It becomes the reverse modified
mutual energy theorem. The media condition is changed to

\begin{equation}
\epsilon(\omega)=-\epsilon^{T}(\omega),\ \ \ \ \ \ \ \ \ \mu(\omega)=-\mu^{T}(\omega)\label{eq:7200-300-2}
\end{equation}
That means the media is anti-symmetric. We do not clear whether or
not this kind media exist in the nature, however if it exist, in this
anti-symmetric media, the above reversed mutual energy theorem is
established. Here we call it ``reversed'' is because this media
is anti-symmetric. It is also used to distinguish it with the theorem
will be discussed in following subsection.

\subsection{The modified mutual energy theorem}

Considering a conjugate transform for the time-reverse transform to
all the variable with subscript ``2'' to the formula Eq.(\ref{eq:7200-250})

\[
[E(\omega),H(\omega),J(\omega),K(\omega),\epsilon(\omega),\mu(\omega)]\equiv r[E_{r}(\omega),H_{r}(\omega),J_{r}(\omega),K_{r}(\omega),\epsilon_{r}(\omega),\mu_{r}(\omega)]
\]
\begin{equation}
=[E_{r}^{*}(\omega),H_{r}^{*}(\omega),J_{r}^{*}(\omega),K_{r}^{*}(\omega),-\epsilon_{r}^{*}(\omega),-\mu_{r}^{*}(\omega)]\label{eq:1000-20-3-1-1}
\end{equation}
There is

\[
-\nabla\cdot(E_{1}(\omega)\times H_{2}^{*}(\omega)+E_{2}^{*}(\omega)\times H_{1}(\omega))
\]

\begin{equation}
=J_{1}(\omega)\cdot E_{2}^{*}(\omega)+E_{1}(\omega)\cdot J_{2}^{*}(\omega)+K_{1}(\omega)\cdot H_{2}^{*}(\omega)+H_{1}(\omega)\cdot K_{2}^{*}(\omega)\label{eq:7200-240-1}
\end{equation}
or

\[
-\intop_{S}(E_{1}(\omega)\times H_{2}^{*}(\omega)+E_{2}^{*}(\omega)\times H_{1}(\omega))\:\hat{n}dS
\]
\begin{equation}
=\intop_{V}(J_{1}(\omega)\cdot E_{2}^{*}(\omega)+E_{1}(\omega)\cdot J_{2}^{*}(\omega)+K_{1}(\omega)\cdot H_{2}^{*}(\omega)+H_{1}(\omega)\cdot K_{2}^{*}(\omega))\, dV\label{eq:7200-250-2-1-1}
\end{equation}
The above can be referred as mutual theorem, or
\begin{equation}
(\xi_{1},\xi_{2})_{\omega}+(\rho_{1},\xi_{2})_{\omega}+(\xi_{1},\rho_{2})_{\omega}=0\label{eq:7200-290-1}
\end{equation}
where

\begin{equation}
(\xi_{1},\xi_{2})_{\omega}=\intop_{S}(E_{1}(\omega)\times H_{2}^{*}(\omega)+E_{2}^{*}(\omega)\times H_{1}(\omega))\:\hat{n}dS\label{eq:7200-260-1}
\end{equation}
\begin{equation}
(\rho_{1},\xi_{2})_{\omega}=\intop_{V}(J_{1}(\omega)\cdot E_{2}^{*}(\omega)+K_{1}(\omega)\cdot H_{2}^{*}(\omega))dV\label{eq:7200-270-1}
\end{equation}
\begin{equation}
(\xi_{1},\rho_{2})_{\omega}=\intop_{V}(E_{1}(\omega)\cdot J_{2}^{*}(\omega)+H_{1}(\omega)\cdot K_{2}^{*}(\omega))dV\label{eq:7200-280-1}
\end{equation}
The media equation becomes,
\begin{equation}
-\epsilon_{2}^{*}(\omega)+\epsilon_{1}^{T}(\omega)=0,\ \ \ \ \ \ \ \ \ -\mu_{2}^{*}(\omega)+\mu_{1}^{T}(\omega)=0\label{eq:7200-300-1}
\end{equation}
or
\begin{equation}
\epsilon_{2}(\omega)=\epsilon_{1}^{\dagger}(\omega),\ \ \ \ \ \ \ \ \ \mu_{2}(\omega)=\mu_{1}^{\dagger}(\omega)\label{eq:7200-300-1-2}
\end{equation}
If $\epsilon_{1}=\epsilon_{2}$ and $\mu_{1}=\mu_{2}$ , the word
``modified'' can be dropped, the above theorem becomes modified
mutual energy theorem. For the modified mutual energy theorem the
media satisfy 
\begin{equation}
\epsilon(\omega)=\epsilon^{\dagger}(\omega),\ \ \ \ \ \ \ \ \ \mu(\omega)=\mu^{\dagger}(\omega)\label{eq:7200-300-1-2-1}
\end{equation}
whech is lossless media. Hence mutual energy theorem is established
in lossless media. The above mutual energy theorem has been derived
by this author in reference\cite{shrzhao1,shrzhao2,shrzhao3}.

In case $\zeta_{1}$ and $\zeta_{2}$ one is retarded potential, and
the other is advanced potential, the surface integral $(\xi_{1},\xi_{2})_{\omega}=0$,
or

\[
\intop_{S}(E_{1}(\omega)\times H_{2}^{*}(\omega)+E_{2}^{*}(\omega)\times H_{1}(\omega))\:\hat{n}dS=0
\]
the proof can been seen from the Appendix 3 . In this case the mutual
energy current will not go to the outside of the the surface.

If there are $N$ electromagnetic fields, the modified mutual energy
theorem is
\begin{equation}
\sum_{i=1,j=1}^{i<j,\, j\leqslant N}((\xi_{i},\xi_{j})_{\omega}+(\rho_{i},\xi_{j})_{\omega}+(\xi_{i},\rho{}_{j})_{\omega})=0\label{eq:2000-100-1}
\end{equation}

\subsection{The modified Lorenz reciprocity theorem}

Considering a conjugate transform corresponding to magnetic mirror
transform Eq.(\ref{eq:1000-20-1-3}) to the above modified mutual
energy theorem to all variable with subscript ``2'', there is

\[
-\nabla\cdot(E_{1}(\omega)\times(-1)H_{2}(\omega)+E_{2}(\omega)\times H_{1}(\omega))
\]

\begin{equation}
=J_{1}(\omega)\cdot E_{2}(\omega)-E_{1}(\omega)\cdot J_{2}(\omega)-K_{1}(\omega)\cdot H_{2}(\omega)+H_{1}(\omega)\cdot K_{2}(\omega)\label{eq:7200-240-1-1}
\end{equation}

\begin{equation}
-\epsilon_{2}(\omega)+\epsilon_{1}^{T}(\omega)=0,\ \ \ \ \ \ \ \ \ -\mu_{2}(\omega)+\mu_{1}^{T}(\omega)=0\label{eq:7200-300-1-1}
\end{equation}
This is modified Lorenz reciprocity theorems. In case $\epsilon_{2}(\omega)=\epsilon_{1}(\omega)=\epsilon$
$\mu_{2}(\omega)=\mu_{1}(\omega)=\mu$, hence
\[
\epsilon^{T}(\omega)=\epsilon(\omega),\ \ \ \ \ \ \ \ \ \mu^{T}(\omega)=\mu(\omega)
\]
The ``modified'' can be dropped. It be comes Lorenz reciprocity
theorem. Lorenz reciprocity theorem is established in symmetric media.

\[
\intop_{S}(E_{1}(\omega)\times H_{2}(\omega)-E_{2}(\omega)\times H_{1}(\omega))\, dt\,\hat{n}dS
\]

\begin{equation}
=\intop_{V}\,(J_{1}(\omega)\cdot E_{2}(\omega)-E_{1}(\omega)\cdot J_{2}(\omega)-K_{1}(\omega)\cdot H_{2}(\omega)+H_{1}(\omega)\cdot K_{2}(\omega)\,\, dV=0\label{eq:4000-140-4}
\end{equation}
In case $\zeta_{1}$ and $\zeta_{2}$ one is retarded potential and
one is advanced potential, we knew from last subsection the surface
integral $(\zeta_{1},\zeta_{2})_{\omega}=0$. Assume $\zeta_{1}$
is retarded potential and $\zeta_{2}$is advanced potential, since
we have did a magnetic mirror transform to $\zeta_{2}$, after the
transform $\zeta_{2}$ become retarded potential. Hence there is if
$\zeta_{1}$ and $\zeta_{2}$ are all retarded potential there is
the surface integral for reciprocity theorem 

\begin{equation}
\intop_{S}(E_{1}(\omega)\times H_{2}(\omega)-E_{2}(\omega)\times H_{1}(\omega))\, dt\,\hat{n}dS=0\label{eq:7200-100-1}
\end{equation}

\subsection{The modified reversed reciprocity theorem}

After a reverse conjugate transform for subscript ``2''. The above
formula can be written as

\begin{equation}
\epsilon_{2}^{*}(\omega)+\epsilon_{1}^{T}(\omega)=0,\ \ \ \ \ \ \ \ \mu_{2}^{*}(\omega)+\mu_{1}^{T}(\omega)=0\label{eq:7400-10}
\end{equation}
and

\[
-\nabla\cdot(-E_{1}(\omega)\times H_{2}^{*}(\omega)+E_{2}^{*}(\omega)\times H_{1}(\omega))
\]
\begin{equation}
=J_{1}(\omega)\cdot E_{2}^{*}(\omega)-J_{2}^{*}(\omega)\cdot E_{1}(\omega)-K_{1}(\omega)\cdot H_{2}^{*}(\omega)+K_{2}^{*}(\omega)\cdot H_{1}(\omega)\label{eq:7400-30}
\end{equation}
The corresponding integral form is

\[
-(E_{1}(\omega)\times H_{2}^{*}(\omega)-E_{2}^{*}(\omega)\times H_{1}(\omega))\:\hat{n}dS
\]
\begin{equation}
=\intop_{V}(J_{1}(\omega)\cdot E_{2}^{*}(\omega)-J_{2}^{*}(\omega)\cdot E_{1}(\omega)-K_{1}(\omega)\cdot H_{2}^{*}(\omega)+K_{2}^{*}(\omega)\cdot H_{1}(\omega))\, dV\label{eq:7400-50}
\end{equation}
This can be referred as the reversed reciprocity theorem. In case
$\epsilon_{2}(\omega)=\epsilon_{1}(\omega)=\epsilon$ $\mu_{2}(\omega)=\mu_{1}(\omega)=\mu$,
the word ``modified'' can be dropped, it becomes reversed reciprocity
theorem. The media condition becomes

\begin{equation}
\epsilon(\omega)=-\epsilon^{\dagger}(\omega),\ \ \ \ \ \ \ \ \mu(\omega)=-\mu^{\dagger}(\omega)\label{eq:7400-10-1}
\end{equation}
This kind media can be referred as anti-lossless media. Hence the
reversed reciprocity theorem is established in anti-lossless media.

\subsection{The surface integral in the mutual energy theorem\label{sub:The-surface-integral-1}}

If both $\zeta_{1}$ and $\zeta_{2}$ are both transmitting fields
or the retarded potential, in general 
\begin{equation}
(\xi_{1},\xi_{2})_{\tau}\neq0\label{eq:450-3-4-1}
\end{equation}
Since, that means $(\xi_{1},\xi_{2})_{\tau}$ are mutual energy current
go through the surface. For example if $\zeta_{1}=\zeta_{2}=\zeta$
then

\begin{equation}
(\xi,\xi)_{\tau}=\intop_{S}\intop_{t=-\infty}^{\infty}(E(t+\tau)\times H(t)+E(t)\times H(t+\tau))\,\hat{n}dS\label{eq:6200-100-1}
\end{equation}
and

\[
F\{(\xi,\xi)_{\tau}\}=\intop_{S}(E(\omega)\times H^{*}(\omega)+E^{*}(\omega)\times H(\omega))\,\hat{n}dS
\]
\begin{equation}
=2\intop_{S}Re\{E(\omega)\times H^{*}(\omega)\}\,\hat{n}dS\label{eq:9000-1}
\end{equation}
$E(\omega)\times H^{*}(\omega)$ is the Fourier domain Poynting vector,
$\intop_{S}Re\{E(\omega)\times H^{*}(\omega)\}\,\hat{n}dS$ is the
power flow out the surface which is not vanish in general. It is only
vanish if the surface $S$ is super conductor or magnetic super conductor
wall. 

Hence there is

\begin{equation}
(\xi,\xi)_{\omega}\neq0\label{eq:6200-200-1}
\end{equation}
and hence, after a inverse Fourier transform, there is
\begin{equation}
(\xi,\xi)_{\tau}=F^{-1}\{(\xi,\xi)_{\omega}\}\neq0\label{eq:6200-210-1}
\end{equation}
in general. If both of them $\zeta_{1}$, $\zeta_{2}$ are the field
of retarded potential, the mutual energy current will have the same
direction from inner side to the outside of the surface $S$, the
surface integral is not vanish in general. 

In other hand if one of them is the field of retarded potential and
the other is the field of advanced potential. For example $\rho_{1}=[J_{1},K_{1}]$
is the source and $\rho_{2}=[J_{2},K_{2}]$ is sink. $\rho_{1}$ and
$\rho_{2}$ are inside the surface $S$. In this case, $\xi_{1}$
is retarded potential. $\xi_{2}$ is advanced potential, there is

\begin{equation}
(\xi_{1},\xi_{2})_{\tau}=0\label{eq:450-5-1-1}
\end{equation}
The proof can been seen in Appendix 3 of the reference{[}22{]}. In
the proof where the Sommerfeld's radiation condition has been applied.

\section{Mutual energy theorems in time domain}

In the section\ref{sec:Derive-the-theorems}, we has derived the mutual
energy theorem directly from Fourier domain. The derivation has first
derived the time-reversed mutual energy theorem from Poynting theorem.
The time reversed transform is applied to derive the mutual energy
theorem from time reversed mutual energy theorem. However time reversed
transform need Maxwell equation to prove. This means actually we did
not derive the mutual energy theorem from Poynting theorem but through
time reversed transform, the Maxwell equation is involved. Hence we
did not purely derived the mutual energy theorem from Poynting theorem.

In this article the mutual energy of a electromagnetic field system
is defined as the difference between the total energy (power) and
the self energy (power). The mutual energy is obtained from the idea
that subtract the self energy from the total energy, the rest is the
mutual energy. There is energy conservation theorem which is Poynting
theorem. The Poynting theorem guarantees that the total energy and
self energy are conservative. Hence the mutual energy should be also
conservative. The mutual energy is conservative is referred as mutual
energy theorem. The following offers the detail of the derivation
of the mutual energy theorem.

\subsection{The instantaneous-time mutual energy theorem }

Assume $\zeta_{1}=[E_{1},H_{1},J_{1},K_{1},D_{1},B_{1}]$ and $\zeta_{2}=[E_{2},H_{2},J_{2},K_{2},D_{2},B_{2}]$
are electromagnetic fields, which can be retarded potential, advanced
potential, time-offset or space-offset. Let $\zeta=\zeta_{1}+\zeta_{2}$
be superimposing electromagnetic field. Assume that $\zeta_{1}$,
$\zeta_{2}$ satisfy Maxwell equation Eq.(\ref{eq:490-1}, \ref{eq:500-1})
and Eq.(\ref{eq:501-1-2-1-1}, \ref{eq:502-1-2-1-1}). Hence $\zeta_{1}$,
$\zeta_{2}$ satisfy the modified Poynting theorem Eq.(\ref{eq:30}),
that means, 
\begin{equation}
-\nabla\cdot(E_{1}\times H_{1})=J_{1}\cdot E_{1}+K_{1}\cdot H_{1}+E_{1}\cdot\partial D_{1}+H_{1}\cdot\partial B_{1}\label{eq:70}
\end{equation}

\begin{equation}
-\nabla\cdot(E_{2}\times H_{2})=J_{2}\cdot E_{2}+K_{2}\cdot H_{2}+E_{2}\cdot\partial D_{2}+H_{2}\cdot\partial B_{2}\label{eq:80}
\end{equation}
Then the superimposing electromagnetic field also satisfies the modified
Poynting theorem Eq.(\ref{eq:30}). 

\[
-\nabla\cdot((E_{1}+E_{2})\times(H_{1}+H_{2}))
\]
\begin{equation}
=(J_{1}+J_{2})\cdot(E_{1}+E_{2})+(K_{1}+K_{2})\cdot(H_{1}+H_{2})+(E_{1}+E_{2})\cdot\partial(D_{1}+D_{2})+(H_{1}+H_{2})\cdot\partial(B_{1}+B_{2})\label{eq:70-1}
\end{equation}
Eq.(\ref{eq:70-1}) tell us the total energy should be conservative.
Eq.(\ref{eq:70},\ref{eq:80}) tell us the self energy is conservative.
Subtract the self energy from the total energy we can obtained the
mutual energy. The mutual energy should be also conservative. Subtract
Eq.(\ref{eq:70},\ref{eq:80}) from Eq.(\ref{eq:70-1}), there is,
\[
-\nabla\cdot(E_{1}\times H_{2}+E_{2}\times H_{1})
\]

\begin{equation}
=J_{1}\cdot E_{2}+J_{2}\cdot E_{1}+K_{1}\cdot H_{2}+K_{2}\cdot H_{1}+E_{1}\cdot\partial D_{2}+E_{2}\cdot\partial D_{1}+H_{1}\cdot\partial B_{2}+H_{2}\cdot\partial B_{1}\label{eq:90}
\end{equation}
The corresponding integral form of the above formula is 
\[
-\intop_{S}(E_{1}\times H_{2}+E_{2}\times H_{1})\cdot\hat{n}dS
\]

\begin{equation}
=\intop_{V}\,(J_{1}\cdot E_{2}+J_{2}\cdot E_{1}+K_{1}\cdot H_{2}+K_{2}\cdot H_{1}+E_{1}\cdot\partial D_{2}+E_{2}\cdot\partial D_{1}+H_{1}\cdot\partial B_{2}+H_{2}\cdot\partial B_{1})\, dV\label{eq:100}
\end{equation}
$V$ is volume, $S$ is the boundary surface of $V$. $S=\partial V$.
Where $\hat{n}$ is norm vector of the surface $S$. The above Eq.(\ref{eq:90},\ref{eq:100})
are referred as the mutual energy theorem. Same as Poynting theorem,
if the medial Eq.(\ref{eq:501-1-2-1-3},\ref{eq:502-1-2-1-3}), The
above formula Eq(\ref{eq:100}) is referred as the mutual energy theorem.
If the medial Eq.(\ref{eq:501-1-2-1-1},\ref{eq:502-1-2-1-1}) is
satisfies, Eq(\ref{eq:100}) is referred as modified mutual energy
theorem. 

Eq(\ref{eq:100}) is too long. Inner product will be defined to shorten
the formula.

\subsection{Inner product of two electromagnetic systems in spatial-temporal
domain}

Assume $\xi_{i}=[E_{i}(t),H_{i}(t)]$, $\eta_{i}=[D_{i}(t),B_{i}(t)]$,
$\rho_{i}=[J_{i}(t),K_{i}(t)]$, $i=1,2$. A inner product on the
surface can be defined as following

\begin{equation}
(\xi_{1}(t),\xi_{2}(t))=\intop_{S}(E_{1}(t)\times H_{2}(t)+E_{2}(t)\times H_{1}(t))\,\hat{n}dS\label{eq:800-1-1}
\end{equation}
In the same way we can also define
\begin{equation}
(\rho_{1}(t),\xi_{2}(t))=\intop_{V}(J_{1}(t)\cdot E_{2}(t)+K_{1}(t)\cdot H_{2}(t))\, dV\label{eq:800-5-3}
\end{equation}
\begin{equation}
(\xi_{1}(t),\partial\eta_{2}(t))=\intop_{V}(E_{1}\cdot\partial D_{2}+H_{1}\cdot\partial B_{2})\, dV\label{eq:800-5-3-1}
\end{equation}
The mutual energy theorem can be rewritten as following,
\begin{equation}
-(\xi_{1},\xi_{2})=(\rho_{1},\xi_{2})+(\xi_{1},\rho_{2})+(\xi_{1},\partial\eta_{2})+(\partial\eta_{1},\xi_{2})\label{eq:450-3-3}
\end{equation}
or
\begin{equation}
(\xi_{1},\xi_{2})+(\rho_{1},\xi_{2})+(\xi_{1},\rho_{2})+(\xi_{1},\partial\eta_{2})+(\partial\eta_{1},\xi_{2})=0\label{eq:450-3-3-3}
\end{equation}
The above formula tell us the summation of the mutual energy current
flow out the surface $S$: $(\xi_{1},\xi_{2})$, the mutual energy
loss contributed from the source $\rho_{1},\rho_{2}$: $(\rho_{1},\xi_{2})+(\xi_{1},\rho_{2})$
and the mutual energy loss in the space: $(\xi_{1},\partial\eta_{2})+(\partial\eta_{1},\xi_{2})$
are zero. The above formula is instantaneous mutual energy theorem.

In case the superimposition electromagnetic field contains $N$ electromagnetic
fields, the above instantaneous mutual energy theorem can be written
as

\begin{equation}
\sum_{i=1,j=1}^{i<j,j\leqslant N}((\xi_{i},\xi_{j})+(\rho_{i},\xi_{j})+(\xi_{i},\rho_{j})+(\xi_{i},\partial\eta_{j})+(\partial\eta_{i},\xi_{j}))=0\label{eq:450-3-3-1}
\end{equation}

\subsection{The modified time-correlated mutual energy theorem}

Considering if we use $\zeta_{1\tau}=\tau\zeta_{1}$ to replace $\zeta_{1}$,
since after the time-offset transform, $\zeta_{1\tau}$ still satisfies
the Maxwell equation, and hence the satisfies the modified Poynting
theorem and the modified mutual energy theorem, hence there is
\[
(\xi_{1}(t+\tau),\xi_{2}(t))+(\rho_{1}(t+\tau),\xi_{2}(t))+(\xi_{1}(t+\tau),\rho_{2}(t))
\]

\begin{equation}
+(\xi_{1}(t+\tau),\partial_{t}\eta_{2}(t))+(\partial_{t+\tau}\eta_{1}(t+\tau),\xi_{2}(t))=0\label{eq:450-3-3-2}
\end{equation}
taking the integral to the above formula to the variable $t$, there
is
\[
\intop_{t=-\infty}^{\infty}((\xi_{1}(t+\tau),\xi_{2}(t))+(\rho_{1}(t+\tau),\xi_{2}(t))+(\xi(t+\tau),\rho_{2}(t))
\]
\begin{equation}
+(\xi_{1}(t+\tau),\partial_{t}\eta_{2}(t))+(\partial_{t+\tau}\eta_{1}(t+\tau),\xi_{2}(t)))\, dt=0\label{eq:450-3-3-2-1}
\end{equation}
In the Appendix 1 it is proven that if the media satisfies
\begin{equation}
\epsilon_{1}^{\dagger}(\omega)=\epsilon_{2}(\omega),\ \ \ \ \mu_{1}^{\dagger}(\omega)=\mu_{2}(\omega)\label{eq:2000-10}
\end{equation}
or after a inverse Fourier transform $F^{-1}\{\bullet\}=\intop_{t=-\infty}^{\infty}\exp(j\omega t)\,\bullet dt$
the media satisfies,
\[
\epsilon_{1}^{T}(-t)=\epsilon_{2}(t),\ \ \ \ \mu_{1}^{T}(-t)=\mu_{2}(t)
\]
The last 2 items of Eq.(\ref{eq:450-3-3-2-1}) disappear i.e., 
\begin{equation}
\intop_{t=-\infty}^{\infty}(\xi_{1}(t+\tau),\partial_{t}\eta_{2}(t))+(\partial_{t+\tau}\eta_{1}(t+\tau),\xi_{2}(t))\, dt=0\label{eq:2000-20}
\end{equation}
Hence there is

\begin{equation}
\intop_{-\infty}^{\infty}(\xi_{1}(t+\tau),\xi_{2}(t))+(\rho_{1}(t+\tau),\xi_{2}(t))+(\xi_{1}(t+\tau),\rho_{2}(t))\, dt=0\label{eq:450-3-3-2-1-1}
\end{equation}
This is referred as the modified time-correlation mutual energy theorem.
In case $\epsilon_{2}(\omega)=\epsilon_{1}(\omega)=\epsilon(\omega)$,
there is 
\begin{equation}
\epsilon^{\dagger}(\omega)=\epsilon(\omega),\ \ \ \ \ \ \mu^{\dagger}(\omega)=\mu(\omega)\label{eq:2000-40}
\end{equation}
or after a inverse Fourier transform
\begin{equation}
\epsilon^{T}(-t)=\epsilon(t)\ \ \ \ \ \ \ \mu^{T}(-t)=\mu(t)\label{eq:2000-50}
\end{equation}
This means the media must be symmetry with time $t$. 

Define new inner product in spatial-temporal space
\begin{equation}
(\xi_{1},\xi_{2})_{\tau}=\intop_{t=-\infty}^{\infty}((\xi_{1}(t+\tau),\xi_{2}^{*}(t))dt\label{eq:2000-60}
\end{equation}
\begin{equation}
(\rho_{1},\xi_{2})_{\tau}=\intop_{t=-\infty}^{\infty}(\rho_{1}(t+\tau),\xi_{2}^{*}(t))dt\label{eq:2000-70}
\end{equation}
\begin{equation}
(\xi_{1},\rho_{2})_{\tau}=\intop_{t=-\infty}^{\infty}(\xi_{1}(t),\rho_{2}^{*}(t+\tau))dt\label{eq:2000-70-1}
\end{equation}
The modified time-correlation mutual energy theorem can be written
as,
\begin{equation}
(\xi_{1},\xi_{2})_{\tau}+(\rho_{1},\xi_{2})_{\tau}+(\xi_{1},\rho_{2})_{\tau}=0\label{eq:2000-80}
\end{equation}

Perhaps you will argue the field variable $\zeta_{1}(t)$ $\zeta_{2}(t)$
are real variable, why the above define the inner product here with
complex conjugate symbol ``{*}'' inside? The reason is in following
subsection we need to make a Fourier transform of above formula that
also need this conjugate symbol.

In case the media satisfies Eq.(\ref{eq:501-1-2-1-3}, \ref{eq:502-1-2-1-3}),
the modified time-correlation mutual energy theorem becomes the time-correlation
mutual energy theorem. The word of ``modified'' is dropped. It also
be referred as time-correlation reciprocity theorem in reference\cite{Adrianus2}.
The above modified time-correlation mutual energy theorem can be written
as following,

\[
\intop_{S}\intop_{t=-\infty}^{\infty}(E_{1}(t+\tau)\times H_{2}^{*}(t)+E_{2}^{*}(t)\times H_{1}(t+\tau))\, dt\,\hat{n}dS
\]
\[
+\intop_{V}\intop_{t=-\infty}^{\infty}(J_{1}(t+\tau)\cdot E_{2}^{*}(t)+K_{1}(t+\tau)\cdot H_{2}^{*}(t)+\rho_{2}(t)+E_{1}(t+\tau)\cdot\rho_{2}^{*}(t)+H_{1}(t+\tau)\cdot K_{2}^{*}(t))\, dt\, dV
\]
\begin{equation}
=0\label{eq:2000-90}
\end{equation}
In the above formula the conjugate symbol ``{*}'' can be dropped
since $\zeta_{1}(t)$ and $\zeta_{2}(t)$ is a real variable, however
we put ``{*}'' there the formula is still correct. It will used
in next subsection. 

If there are $N$ electromagnetic fields, the modified time-correlation
mutual energy theorem is
\begin{equation}
\sum_{i=1,j=1}^{i<j,\, j\leqslant N}((\xi_{i},\xi_{j})_{\tau}+(\rho_{i},\xi_{j})_{\tau}+(\xi_{i},\rho{}_{j})_{\tau})=0\label{eq:2000-100}
\end{equation}

\section{Mutual energy theorems in Fourier domain}

\subsection{The modified complex mutual energy theorem}

Considering $F\{\bullet\}=\intop_{t=-\infty}^{\infty}\exp(-j\omega t)\,\bullet dt$
is Fourier transform, we have 
\[
(\xi_{1},\xi_{2})_{\omega}\equiv F\{(\xi_{1},\xi_{2})_{\tau}\}
\]

\begin{equation}
=\intop_{S}(E_{1}(\omega)\times H_{2}^{*}(\omega)+E_{2}^{*}(\omega)\times H_{1}(\omega))\cdot\hat{n}dS\label{eq:2000-110}
\end{equation}
Please see the Appendix 1 for definition of $F\{\bullet\}$ and the
details of calculation.
\[
(\rho_{1},\xi_{2})_{\omega}\equiv F\{(\rho_{1},\xi_{2})_{\tau}\}
\]
\begin{equation}
=\intop_{V}(J_{1}(\omega)\cdot E_{2}^{*}(\omega)+K_{1}(\omega)\cdot H_{2}^{*}(\omega))\, dV\label{eq:2000-120}
\end{equation}
\[
(\xi_{1},\rho_{2})_{\omega}\equiv F\{(\xi_{1},\rho_{2},)_{\tau}\}
\]
\begin{equation}
=\intop_{V}(E_{1}(\omega)\cdot J_{2}^{*}(\omega)+H_{1}(\omega)\cdot K_{2}^{*}(\omega))\, dV\label{eq:2000-120-1}
\end{equation}
the corresponding to frequency theorem is 
\begin{equation}
(\xi_{1},\xi_{2})_{\omega}+(\rho_{1},\xi_{2})_{\omega}+(\xi_{1},\rho_{2})_{\omega}=0\label{eq:2000-130}
\end{equation}
or
\[
\intop_{S}(E_{1}(\omega)\times H_{2}^{*}(\omega)+E_{2}^{*}(\omega)\times H_{1}(\omega))\cdot\hat{n}dS
\]
\[
+\intop_{V}(J_{1}(\omega)\cdot E_{2}^{*}(\omega)+E_{1}(\omega)\cdot J_{2}^{*}(\omega)+K_{1}(\omega)\cdot H_{2}^{*}(\omega)+H_{1}(\omega)K_{2}^{*}(\omega))\, dV
\]
\begin{equation}
=0\label{eq:2000-140}
\end{equation}
Where the media have to meet the condition,
\begin{equation}
\epsilon_{1}^{\dagger}(\omega)=\epsilon_{2}(\omega),\ \ \ \ \mu_{1}^{\dagger}(\omega)=\mu_{2}(\omega)\label{eq:2000-150}
\end{equation}
This formula has been referred the modified complex mutual energy
theorem\cite{shrzhao1,shrzhao2,shrzhao3} by this author. In the reference
\cite{shrzhao1,shrzhao2,shrzhao3}, This author has obtained the modified
complex mutual energy theorem from modified reciprocity theorem through
a conjugate transform. Conjugate transform is the magnetic mirror
transform in Fourier domain, see sub-section 2.5. In this article,
the above complex mutual energy theorem is re-obtained through the
modified Poynting theorem and the concept of mutual energy.

If the $\zeta_{1}$ and $\zeta_{2}$ are in the same media
\begin{equation}
\epsilon(\omega)=\epsilon_{1}(\omega)=\epsilon_{2}(\omega)\ \ \ \ \mu(\omega)=\mu_{1}(\omega)=\mu_{2}(\omega)\label{eq:2000-160}
\end{equation}
There is
\begin{equation}
\epsilon^{\dagger}(\omega)=\epsilon(\omega),\ \ \ \ \mu^{\dagger}(\omega)=\mu(\omega)\label{eq:2000-170}
\end{equation}
This is referred as lossless media. In lossless media the corresponding
theorem is referred as the complex mutual energy theorem. The word
``modified'' can be dropped, if $\epsilon_{1}(\omega)=\epsilon_{2}(\omega)$
and $\mu_{1}(\omega)=\mu_{2}(\omega)$. The complex mutual energy
theorem has been rediscovered later and is referred as the second
reciprocity theorem\cite{Petrusenko}. 

If there are $N$ electromagnetic fields, the corresponding mutual
energy theorem, 
\begin{equation}
\sum_{i=1,j=1}^{i<j,\, j\leqslant N}((\xi_{i},\xi_{j})_{\omega}+(\rho_{i},\xi_{j})_{\omega}+(\xi_{i},\rho{}_{j})_{\omega})=0\label{eq:2000-180}
\end{equation}

\subsection{The surface integral in the mutual energy theorem\label{sub:The-surface-integral}}

If both $\zeta_{1}$ and $\zeta_{2}$ are both retarded potential,
in general 
\begin{equation}
(\xi_{1},\xi_{2})_{\tau}\neq0\label{eq:450-3-4}
\end{equation}
Since, that means $(\xi_{1},\xi_{2})_{\tau}$ are mutual energy current
go through the surface. For example if $\zeta_{1}=\zeta_{2}=\zeta$
then

\begin{equation}
(\xi,\xi)_{\tau}=\intop_{S}\intop_{t=-\infty}^{\infty}(E(t+\tau)\times H(t)+E(t)\times H(t+\tau))\,\hat{n}dS\label{eq:6200-100}
\end{equation}
and

\[
F\{(\xi,\xi)_{\tau}\}=\intop_{S}(E(\omega)\times H^{*}(\omega)+E^{*}(\omega)\times H(\omega))\,\hat{n}dS
\]
\begin{equation}
=2\intop_{S}Re\{E(\omega)\times H^{*}(\omega)\}\,\hat{n}dS\label{eq:9000}
\end{equation}
$E(\omega)\times H^{*}(\omega)$ is the Fourier domain Poynting vector,
$\intop_{S}Re\{E(\omega)\times H^{*}(\omega)\}\,\hat{n}dS$ is the
power flow out the surface which is not vanish in general. It is only
vanish if the surface $S$ is super conductor or magnetic super conductor
wall. 

Hence there is

\begin{equation}
(\xi,\xi)_{\omega}\neq0\label{eq:6200-200}
\end{equation}
and hence, after a inverse Fourier transform, there is
\begin{equation}
(\xi,\xi)_{\tau}=F^{-1}\{(\xi,\xi)_{\omega}\}\neq0\label{eq:6200-210}
\end{equation}
in general. If both of them $\zeta_{1}$, $\zeta_{2}$ are the field
of retarded potential, the mutual energy current will have the same
direction from inner side to the outside of the surface $S$, the
surface integral is not vanish in general. 

In other hand if one of them is the field of retarded potential and
the other is the field of advanced potential. For example $\rho_{1}=[J_{1},K_{1}]$
is the source and $\rho_{2}=[J_{2},K_{2}]$ is sink. $\rho_{1}$ and
$\rho_{2}$ are inside the surface $S$. In this case, $\xi_{1}$
is retarded potential. $\xi_{2}$ is advanced potential, there is

\begin{equation}
(\xi_{1},\xi_{2})_{\tau}=0\label{eq:450-5-1}
\end{equation}
The proof can been seen in Appendix 3. In the proof where the Sommerfeld's
radiation condition has been applied.

\section{Reciprocity theorems}

In this section we assume the mutual energy theorem is known but the
reciprocity theorem is unknown. We derive the reciprocity theorem
from the mutual energy theorem. This way we show that the reciprocity
theorem actually is a sub-theorem of mutual energy theorem. Actually
is it is a special situation of the mutual energy theorem.

\subsection{Time convolution reciprocity theorem}

In above mutual energy theorems, $\zeta_{1}$ and $\zeta_{2}$ can
be retarded potential or advanced potential or even the combination
of retarded potential and advanced potential. In a special situation
where $\zeta_{1}$ and $\zeta_{2}$ one is retarded potential and
the other one is advanced potential. Assume $\zeta_{1}$ is the retarded
potential, $\zeta_{2}$ is the advanced potential . From the above
case we have know in mutual energy theorem the surface integral vanish
on the infinite sphere $S$.
\begin{equation}
(\xi_{1},\xi_{2})_{\tau}=0\label{eq:4000-10}
\end{equation}
or
\begin{equation}
\intop_{S}\intop_{t=-\infty}^{\infty}(E_{1}(t+\tau)\times H_{2}(t)+E_{2}(t)\times H_{1}(t+\tau)\, dt\,\hat{n}dS=0\label{eq:4000-20}
\end{equation}
Hence according the time correlation mutual energy theorem Eq(\ref{eq:2000-80}),
there is 
\begin{equation}
(\rho_{1},\xi_{2})_{\tau}+(\xi_{1},\rho_{2})_{\tau}=0\label{eq:4000-50}
\end{equation}
or

\begin{equation}
\intop_{V}\,\intop_{-\infty}^{\infty}(J_{1}(t+\tau)\cdot E_{2}(t)+E_{1}(t+\tau)\cdot J_{2}(t)+K_{1}(t+\tau)\cdot H_{2}(t)+H_{1}(t+\tau)\cdot K_{2}(t)\, dt\, dV=0\label{eq:4000-60}
\end{equation}
Since we know $\zeta_{2}$ is advanced potential, hence $\zeta_{h2}=h\zeta_{2}$
become retarded potential. Here $h$ is magnetic mirror transform
defined in Eq.(\ref{eq:1000-20-1}). Hence $\zeta_{2}=h\zeta_{h2}$,
or
\[
[E_{2}(t),H_{2}(t),J_{2}(t),K_{2}(t),\epsilon_{2}(t),\mu_{2}(t)]
\]
 
\begin{equation}
=[E_{h2}(-t),-H_{h2}(-t),-J_{h2}(-t),K_{h2}(-t),\epsilon_{h2}(-t),\mu_{h2}(-t)]\label{eq:20000-10}
\end{equation}
or substitute this to the above formula Eq.(\ref{eq:4000-20},\ref{eq:4000-60})
there is

\begin{equation}
\intop_{S}\intop_{t=-\infty}^{\infty}(E_{1}(t+\tau)\times(-1)H_{h2}(-t)+E_{h2}(-t)\times H_{1}(t+\tau))\, dt\,\hat{n}dS=0\label{eq:4000-70-1}
\end{equation}
and
\[
\intop_{V}\,\intop_{-\infty}^{\infty}(J_{1}(t+\tau)\cdot E_{h2}(-t)+(-1)J_{h2}(-t)\cdot E_{1}(t+\tau)+K_{1}(t+\tau)\cdot(-1)H_{h2}(-t)+K_{h2}(-t)\cdot H_{1}(t+\tau)\, dt\, dV
\]

\begin{equation}
=0\label{eq:4000-80-1}
\end{equation}
or in the above integral substitute $t'=-t$

\begin{equation}
\intop_{S}\intop_{t'=-\infty}^{\infty}(E_{1}(-t'+\tau)\times(-1)H_{h2}(t')+E_{h2}(t')\times H_{1}(-t'+\tau))\, dt'\,\hat{n}dS=0\label{eq:4000-70-1-1}
\end{equation}
and
\[
\intop_{V}\,\intop_{t'=-\infty}^{\infty}(J_{1}(-t'+\tau)\cdot E_{h2}(t')+(-1)J_{h2}(t')\cdot E_{1}(-t'+\tau)
\]

\begin{equation}
+K_{1}(-t'+\tau)\cdot(-1)H_{h2}(t)+K_{h2}(t')\cdot H_{1}(-t'+\tau)\, dt'\, dV=0\label{eq:4000-80-1-1}
\end{equation}
using $t$ to replace $t'$, there is 

\begin{equation}
\intop_{S}\intop_{t'=-\infty}^{\infty}(-E_{1}(\tau-t)\times H_{h2}(t)+E_{h2}(t)\times H_{1}(\tau-t))\, dt\,\hat{n}dS=0\label{eq:4000-70-1-1-1}
\end{equation}
and
\[
\intop_{V}\,\intop_{t'=-\infty}^{\infty}(J_{1}(\tau-t)\cdot E_{h2}(t)-J_{h2}(t)\cdot E_{1}(\tau-t)-K_{1}(\tau-t)\cdot H_{h2}(t)+K_{h2}(t)\cdot H_{1}(\tau-t)\, dt\, dV
\]

\begin{equation}
=0\label{eq:4000-80-1-1-1}
\end{equation}

The media formula 

\begin{equation}
\epsilon_{1}^{T}(-t)=\epsilon_{2}(t),\ \ \ \ \mu_{1}^{T}(-t)=\mu_{2}(t)\label{eq:7000-300}
\end{equation}
after the substitution Eq.(\ref{eq:20000-10}) become
\begin{equation}
\epsilon_{1}^{T}(-t)=\epsilon_{h2}(-t),\ \ \ \ \mu_{1}^{T}(-t)=\mu_{h2}(-t)\label{eq:7100-310}
\end{equation}
or
\begin{equation}
\epsilon_{1}^{T}(t)=\epsilon_{h2}(t),\ \ \ \ \mu_{1}^{T}(t)=\mu_{h2}(t)\label{eq:7100-320}
\end{equation}
Considering $\zeta_{h2}$ is retarded potential. We can take the subscript
``$h$'' and keep in mind that $\zeta_{2}$is the retarded potential,
there is,

\begin{equation}
\intop_{S}\intop_{t=-\infty}^{\infty}(-E_{1}(\tau-t)\times H_{2}(t)+E_{2}(t)\times H_{1}(\tau-t))\, dt\,\hat{n}dS=0\label{eq:4000-90}
\end{equation}
and

\begin{equation}
\intop_{V}\,\intop_{-\infty}^{\infty}(J_{1}(\tau-t)\cdot E_{2}(t)-E_{1}(\tau-t)\cdot J_{2}(t)-K_{1}(\tau-t)\cdot H_{2}(t)+H_{1}(\tau-t)\cdot K_{2}(t)\, dt\, dV=0\label{eq:4000-100}
\end{equation}
In the above formula the item of the surface integral is zero is only
correct for this case where $\zeta_{1}$ and $\zeta_{2}$ are all
retarded potential (or all are advanced potential). In general case
the surface integral are not zero. Hence in the above formula the
surface integral is still put there. The media should satisfy 
\begin{equation}
\epsilon_{1}^{T}(t)=\epsilon_{2}(t),\ \ \ \ \mu_{1}^{T}(t)=\mu_{2}(t)\label{eq:7100-400}
\end{equation}
The above last second formula can be rewritten as

\[
\intop_{V}\,\intop_{-\infty}^{\infty}(J_{1}(\tau-t)\cdot E_{2}(\tau)-K_{1}(\tau-t)\cdot H_{2}(\tau))\, dt\, dV
\]
\begin{equation}
=\intop_{V}\,\intop_{-\infty}^{\infty}(J_{2}(\tau)\cdot E_{1}(\tau-t)-K_{2}(\tau)\cdot H_{1}(\tau-t))\, dt\, dV\label{eq:4000-100-1}
\end{equation}
In the above formula we keep in mind that both $\zeta_{1}$ $\zeta_{2}$
are all retarded potential. This is the modified convolution reciprocity
theorem\cite{Welch,Adrianus}. In the modified convolution reciprocity
theorem, the media can be arbitrary, it does not need to be symmetry.
However if the media is symmetry, We can choose $\epsilon_{1}=\epsilon_{2}=\epsilon$,
$\mu_{1}=\mu_{2}=\mu$, 
\begin{equation}
\epsilon^{T}(t)=\epsilon(t),\ \ \ \ \mu^{T}(t)=\mu(t)\label{eq:4000-130-1}
\end{equation}
In this situation, ``modified'' can be dropped. So the modified
convolution reciprocity becomes the convolution reciprocity theorem.

\subsection{Lorenz reciprocity theorem}

Assume $\zeta_{1}$ and $\zeta_{2}$ are retarded potential. Considering
the Fourier transform of the above time-convolution reciprocity theorem,
there is
\[
\intop_{S}(-E_{1}(\omega)\times H_{2}(\omega)+E_{2}(\omega)\times H_{1}(\omega))\, dt\,\hat{n}dS+
\]

\begin{equation}
\intop_{V}\,(J_{1}(\omega)\cdot E_{2}(\omega)-E_{1}(\omega)\cdot J_{2}(\omega)-K_{1}(\omega)\cdot H_{2}(\omega)+H_{1}(\omega)\cdot K_{2}(\omega)\,\, dV=0\label{eq:4000-140}
\end{equation}
in case , $\zeta_{1}$ and $\zeta_{2}$ are retarded potential, there
is 
\begin{equation}
\intop_{S}(-E_{1}(\omega)\times H_{2}(\omega)+E_{2}(\omega)\times H_{1}(\omega))\, dt\,\hat{n}dS=0\label{eq:7200-100}
\end{equation}
The above last second formula can be rewritten as
\[
\intop_{V}\,(J_{1}(\omega)\cdot E_{2}(\omega)-K_{1}(\omega)\cdot H_{2}(\omega))\,\, dV
\]
\begin{equation}
=\intop_{V}(J_{2}(\omega)\cdot E_{1}(\omega)-H_{1}(\omega)\cdot K_{2}(\omega))\,\, dV\label{eq:4000-140-1}
\end{equation}
After the Fourier transform the media condition become,
\begin{equation}
\epsilon_{1}^{T}(\omega)=\epsilon_{2}(\omega),\ \ \ \ \mu_{1}^{T}(\omega)=\mu_{2}(\omega)\label{eq:4000-150}
\end{equation}
This is the modified reciprocity theorem\cite{J_A_Kong2}. In case
choose $\epsilon_{1}=\epsilon_{2}=\epsilon$, $\mu_{1}=\mu_{2}=\mu$,
and hence have,

\begin{equation}
\epsilon^{T}(\omega)=\epsilon(\omega),\ \ \ \ \mu^{T}(\omega)=\mu(\omega)\label{eq:4000-160}
\end{equation}
The ``modified'' can be dropped, it become the reciprocity theorem,
or the Lorentz reciprocity theorem\cite{Rumsey}. Lorenz reciprocity
theorem can be obtained also thorough conjugate transform from the
Fourier domain mutual energy theorem.

According to the above discussion the Lorentz reciprocity theorem,
and time-convolution reciprocity theorem are a special case of the
mutual energy theorem where the two electromagnetic fields one is
the retarded potential and the other one is the advanced potential. 

In the Lorenz reciprocity theorem and convolution reciprocity theorem
the two fields are both retarded potentials, even originally one is
retarded potential and another is advanced potential in the time-correlation
mutual energy theorem or complex mutual energy theorem. The concept
reaction\cite{Rumsey_VH} is a special mutual energy (power) where
two fields are in opposite, one is retarded potential the other one
is the advanced potential.

\subsection{The relationship of Poynting theorem, mutual energy theorem and reciprocity
theorem}

Even the modified complex mutual energy theorem and modified Lorenz
reciprocity theorem can be derived from each other through a conjugate
transform, the modified time-correlation mutual energy theorem and
time-convolution reciprocity theorem can be derived through a magnetic
mirrored transform, they are still independent theorems. The reason
is that the mirror transform and conjugate transform are not mathematical
equation, it is physical equation which contains some information
coming from the Maxwell equation. When conjugate transform or mirror
transform is applied in a derivation, it is same as the Maxwell equation
is used again. 

If we drop out the word ``modified'', The complex mutual energy
theorem and the Lorenz reciprocity theorem are thoroughly different
theorems, the complex mutual energy theorem is established in losseless
media and the Lorenz reciprocity theorem is established in symmetry
media. They are suitable in different situation and are different
theorems. 

However even the word ``modified'' is dropped, the time-correlation
mutual energy theorem and the complex mutual energy theorem can be
derived from Poynting theorem. This can be proved exactly following
the derivation of this article but do not use the word ``modified''.
Hence time-correlation mutual energy theorem and complex mutual energy
theorem are really a sub-theorem of Poynting theorem. From this point
of view, time-correlation mutual energy theorem and complex mutual
energy theorem (with and without ``modified'') are much closer related
to the Poynting theorem than the convolution reciprocity theorem and
the Lorenz reciprocity theorem.

Time-correlation mutual energy theorem and complex mutual energy theorem
can be easily extended to the situation there is $N$ fields. In principle,
the reciprocity theorem can do the same, however if it is done, there
will be too many minus and positive sign in the extended theorem which
will confuse all of us.

\section{The application of mutual energy theorem}

\subsection{Inner product}

The 3 inner product $(\xi_{1},\xi_{2})$, $(\xi_{1},\xi_{2})_{\tau=0}$,
$(\xi_{1},\xi_{2})_{\omega}$ have been defined in Eq.(\ref{eq:800-1-1},\ref{eq:2000-60}
and \ref{eq:2000-110}), it should be remarkable these are not just
a notation for simplification. These 3 inner products are the real
inner products. It can be proved that these inner products satisfy
the inner product standard 3 conditions as following, if the electromagnetic
fields $\zeta_{1}$ and $\zeta_{2}$ are all retarded potential, there
are

1. Positive-definiteness:

\begin{equation}
(\xi,\xi)\geqq0\,\ \ \ (\xi,\xi)=0\ iff\ \ \xi=0\label{eq:800-2-1}
\end{equation}

2. Conjugate symmetry:
\begin{equation}
(\xi_{1},\xi_{2})=(\xi_{2},\xi_{1})^{*}\label{eq:800-3-1}
\end{equation}

or if $(\xi_{2},\xi_{1})$ is real,

\begin{equation}
(\xi_{1},\xi_{2})=(\xi_{2},\xi_{1})\label{eq:800-3-1-1}
\end{equation}

3. Linearity:
\begin{equation}
(a\xi_{1}+b\xi_{2},\xi_{3})=a(\xi_{1},\xi_{3})+b(\xi_{2},\xi_{3})\label{eq:800-4-1}
\end{equation}
Where $a$ and $b$ are any constant. Here $(\xi_{1},\xi_{2})$ represent
all the 3 inner products $(\xi_{1},\xi_{2})$, $(\xi_{1},\xi_{2})_{\tau=0}$
and $(\xi_{1},\xi_{2})_{\omega}$. 

With the inner product, the norm can be defined as

\begin{equation}
||\xi||=\sqrt{(\xi,\xi)}\label{eq:800-5-1}
\end{equation}

Using the inner product, the mutual energy theorem in the Fourier
domain and in time domain has nearly same formula, the only difference
is the subscript of $\omega$ or $\tau$. 

It is worth to notice that $(\xi_{1},\xi_{2})_{\tau}$ does not satisfy
the above standard inner product conditions.
\[
(\xi_{1},\xi_{2})_{\tau}=\intop_{S}\intop_{t=-\infty}^{\infty}(E_{1}(t+\tau)\times H_{2}^{*}(t)+E_{2}^{*}(t)\times H_{1}(t+\tau))\, dt\,\hat{n}dS
\]
\[
=\intop_{S}\intop_{t=-\infty}^{\infty}(E_{1}(t')\times H_{2}^{*}(t'-\tau)+E_{2}^{*}(t'-\tau)\times H_{1}(t'))\, dt\,\hat{n}dS
\]
\[
=\intop_{S}\intop_{t=-\infty}^{\infty}(E_{2}^{*}(t'-\tau)\times H_{1}(t')+E_{1}(t')\times H_{2}^{*}(t'-\tau))\, dt\,\hat{n}dS
\]
\[
=(\intop_{S}\intop_{t=-\infty}^{\infty}(E_{2}(t'-\tau)\times H_{1}^{*}(t')+E_{1}^{*}(t')\times H_{2}(t'-\tau))\, dt\,\hat{n}dS)^{*}
\]

\begin{equation}
=(\xi_{2},\xi_{1})_{-\tau}^{*}\label{eq:11100-100}
\end{equation}
However if let $\tau=0$, the above formula means that
\begin{equation}
(\xi_{1},\xi_{2})_{\tau=0}=(\xi_{2},\xi_{1})_{\tau=0}^{*}\label{eq:11100-110}
\end{equation}
Hence $(\xi_{1},\xi_{2})_{\tau=0}$ is a good inner product.

\subsection{Applied the mutual energy theorem to the wave expansion}

Assume $\zeta=(E,H,J,K,\epsilon,\mu)$ is a field of retarded potential.
$\zeta$ is in spatial-temporal domain or in Fourier domain. Choose
that $\zeta_{i}=(E_{i},H_{i},J_{i},K_{i},\epsilon_{i},\mu_{i})$ as
also retarded potential. $\zeta_{i}$ is at the same domain as $\zeta$.
$i=0,1,\cdots\infty$. It can be taken that $\epsilon_{i}(\omega)=\epsilon_{0}(\omega)=\epsilon^{\dagger}(\omega)$,
$\mu_{i}(\omega)=\mu_{0}(\omega)=\mu^{\dagger}(\omega)$ in complex
space, or time space $\epsilon_{i}(\tau)=\epsilon_{0}(\tau)=\epsilon^{T}(-\tau)$,
$\mu_{i}(\tau)=\mu_{0}(\tau)=\mu^{T}(-\tau)$. Here $\epsilon_{0}$
and $\mu_{0}$ are not the permittivity and permeability in empty
space, instead, they are permittivity and permeability of the electromagnetic
field $\zeta_{i}$ when $i=0$. Hence there is $\zeta_{i}=[E_{i},H_{i},J_{i},H_{i},\epsilon_{0},\mu_{0}]$.
In the following the inner product $(\xi,\xi_{i})$ also means ether
in time domain which is $(\xi,\xi_{i})_{\tau=0}$ or in Fourier domain
which is $(\xi,\xi_{i})_{\omega}$. It can be in complex space or
in spatial-temporal space. If there is any method the electromagnetic
field $\xi$ can be written as a expansion form
\begin{equation}
\xi=\sum_{i=0}^{\infty}a_{i}\xi_{i}\label{eq:451-2}
\end{equation}
Where $\xi=[E(x,\tau),H(x,\tau)]$ or $\xi=[E(x,\omega),H(x,\omega)]$,
here $x=[x_{1},x_{2},x_{3}]$ is a space variable which is often does
not write out. $x$ can be express according other coordinates for
example spherical coordinates. $a_{i}$ is expansion coefficients
which need to be found in the following. 
\begin{equation}
\xi_{i}(x,\omega)=R_{l}(r)Y_{mm}(\theta,\phi)\label{eq:11200-10}
\end{equation}
$(r,\theta,\phi)$ are spherical coordinates, $Y_{nm}(\theta,\phi)$
is a orthogonal function on the for $\theta$ and $\phi$ variable.
$R_{n}(r)$ is a orthogonal variable alone the variable $r$, and
$(r,\theta,\phi)$ is the spherical coordinates. The index $i=[l,m,n]$.
\begin{equation}
\xi_{i}(x,t)=R_{l}(r)Y_{mn}(\theta,\phi)\varPhi_{k}(t)\label{eq:11200-20}
\end{equation}
where $\varPhi_{k}(t)$ is the Fourier series,
\begin{equation}
\varPhi_{k}(t)=\exp(j\frac{2k\pi t}{P})\label{eq:11200-30}
\end{equation}
The Fourier series is expansion in the region $-\frac{P}{2}\leq t\leq\frac{P}{2}$.
In the numerical calculation for the time variable a fixed number
is used to replace $-\infty<t<\infty$. The index $i=[l,m,n,k]$.

Assume $\xi_{i}$ is with the property of normalized orthogonality,
\begin{equation}
(\xi_{i},\xi_{j})=\delta_{ij}\label{eq:452-1}
\end{equation}
where $(\xi_{i},\xi_{j})$ means $(\xi_{i},\xi_{j})_{\tau=0}$ or
$(\xi_{i},\xi_{j})_{\omega}$ and 
\begin{equation}
\delta_{ij}=\begin{cases}
\begin{array}{c}
1\\
0
\end{array} & \begin{array}{c}
if\ \ i=j\\
if\ \ i\neq j
\end{array}\end{cases}\label{eq:453-1}
\end{equation}
considering Eq.(\ref{eq:451-2}) with \ref{eq:452-1}), there is 
\begin{equation}
(\xi,\xi_{i})=a_{i}\label{eq:454-1}
\end{equation}
where from the modified mutual energy theorem Eq.(\ref{eq:2000-80}
or \ref{eq:2000-130}) we know that
\begin{equation}
(\xi,\xi_{i})=-(\rho,\xi_{i})-(\xi,\rho_{i})\label{eq:455-1}
\end{equation}
or the expansion can be written as 
\begin{equation}
\xi=-\sum_{i=0}^{\infty}((\rho,\xi_{i})+(\xi,\rho_{i}))\xi_{i}\label{eq:456-1}
\end{equation}
In general if both $\xi$ and $\xi_{i}$ are both retarded potential
or both are advanced potential the inner product does not disappear.
Hence the electromagnetic field $\xi$ can be expanded as $\xi_{i}$.
It is worth to see the above expansion can be done in Fourier domain
and also in time domain. The medial $\epsilon(\omega),\mu(\omega)$
can be arbitrary. $\epsilon(\omega),\mu(\omega)$ do not need to be
lossless, because $\epsilon_{0},\mu_{0}$ can always be chosen so
to satisfy Eq.(\ref{eq:2000-150}) even with loss media $\epsilon(\omega),\mu(\omega)$.

The spherical wave expansion and plane wave expansion in Fourier domain
can be found in reference\cite{shrzhao1,shrzhao3}. Where the modified
mutual energy theorem is applied in Fourier domain. In this article
the expansion method has been extended to the time domain. Similar
discussions about the wave expansion can be found also in reference\cite{A_A_Barybin,D_Marcuse}.

\subsection{One example of mutual energy theorem}

Assume there are electromagnetic field systems $\zeta_{1}$,$\zeta_{2}$
and $\zeta_{3}$ are known. Assume, $\zeta_{1}$,$\zeta_{2}$ are
retarded potential, $\zeta_{3}$ is an advanced potential. Please
find out the mutual energy current radiate to the outside of the infinite
sphere $S$.

Solution: the all mutual energy current radiate to the out of the
surface $S$ is $\sum_{i=1,j=1}^{i<j,j\leqslant3}(\xi_{i},\xi_{j})$

Since $\zeta_{3}$ is advanced potential, and $\zeta_{1}$ and $\zeta_{2}$
are retarded potential, there is 
\begin{equation}
(\xi_{1},\xi_{3})=0,\ \ \ \ (\xi_{2},\xi_{3})=0\label{eq:450-1-1-13}
\end{equation}
The mutual energy current radiate to the outside of the surface $S$
is
\[
\sum_{i=1,j=1}^{i<j,j\leqslant3}(\xi_{i},\xi_{j})=(\xi_{1},\xi_{2})+(\xi_{1},\xi_{3})+(\xi_{2},\xi_{3})
\]
\begin{equation}
=(\xi_{1},\xi_{2})=-((\rho_{1},\xi_{2})+(\xi_{1},\rho_{2}))\label{eq:450-1-1-14}
\end{equation}
In the last step the mutual energy theorem for $\zeta_{1}$ and $\zeta_{2}$
has been applied. $\zeta_{1}$ and $\zeta_{2}$ are retarded potential,
$(\xi_{1},\xi_{2})\neq0$. Finished. 

In this case the result is only the mutual energy current of $\zeta_{1}$
and $\zeta_{2}$ which have the contribution to the energy current
going to the outside of the surface $S$.

\section{Complementary theorems}

Chen-To Tai has derived the complementary reciprocity theorem\cite{Chen-to_tai}.
We have obtained 4 theorems 2 mutual energy theorem and 2 reciprocity
theorem. We apply the electromagnetic field swapping transform 

\[
\zeta_{s}=s\zeta=[ZH,\frac{1}{Z}E,-\frac{1}{Z}K,-ZJ,-\frac{1}{Z^{2}}\mu,-Z^{2}\epsilon]
\]
to the 4 theorem, 4 complementary theorems can be obtained, among
them one is the Chen-to Ta's complementary theorem. Consider swapping
transform for the following 4 theorems

\subsection{Corresponding to Lorenz reciprocity theorem}

\[
\intop_{S}(E_{1}(\omega)\times H_{2}(\omega)-E_{2}(\omega)\times H_{1}(\omega))\, dt\,\hat{n}dS
\]

\begin{equation}
=\intop_{V}\,(J_{1}(\omega)\cdot E_{2}(\omega)-E_{1}(\omega)\cdot J_{2}(\omega)-K_{1}(\omega)\cdot H_{2}(\omega)+H_{1}(\omega)\cdot K_{2}(\omega)\,\, dV\label{eq:4000-140-4-1}
\end{equation}
\begin{equation}
\epsilon_{1}(\omega)=\epsilon_{2}^{T}(\omega),\ \ \ \ \ \ \ \ \ \mu_{1}(\omega)=\mu_{2}^{T}(\omega)\label{eq:7200-300-1-1-1}
\end{equation}

\[
\zeta_{s}=s\zeta=[ZH,\frac{1}{Z}E,-\frac{1}{Z}K,-ZJ,-\frac{1}{Z^{2}}\mu,-Z^{2}\epsilon]
\]
The corresponding theorem is

\[
\intop_{S}(E_{1}(\omega)\times\frac{1}{Z}E_{2}(\omega)-ZH_{2}(\omega)\times H_{1}(\omega))\, dt\,\hat{n}dS
\]

\begin{equation}
=\intop_{V}\,(J_{1}(\omega)\cdot ZH_{2}(\omega)-E_{1}(\omega)\cdot(-\frac{1}{Z}K_{2}(\omega))-K_{1}(\omega)\cdot\frac{1}{Z}E_{2}(\omega)+H_{1}(\omega)\cdot(-ZJ_{2}(\omega))\,\, dV\label{eq:4000-140-4-1-1}
\end{equation}
or

\[
\intop_{S}(E_{1}(\omega)\times E_{2}(\omega)-Z^{2}H_{2}(\omega)\times H_{1}(\omega))\, dt\,\hat{n}dS
\]

\begin{equation}
=\intop_{V}\,(Z^{2}J_{1}(\omega)\cdot H_{2}(\omega)+E_{1}(\omega)\cdot K_{2}(\omega)-K_{1}(\omega)\cdot E_{2}(\omega)-Z^{2}H_{1}(\omega)\cdot J_{2}(\omega))\,\, dV\label{eq:4000-140-4-1-1-1}
\end{equation}
\begin{equation}
\epsilon_{1}(\omega)=-\frac{1}{Z^{2}}\mu_{2}^{T}(\omega),\ \ \ \ \ \ \ \ \ \mu_{1}(\omega)=-Z^{2}\epsilon_{2}^{T}(\omega)\label{eq:7200-300-1-1-1-1}
\end{equation}
This is complementary reciprocity of Chen-To Ta. In the above derivation
we have applied the concept of ``modified'', if take the ``modified''
away, there is,

\begin{equation}
\epsilon(\omega)=-\frac{1}{Z^{2}}\mu^{T}(\omega),\ \ \ \ \ \ \ \ \ \mu(\omega)=-Z^{2}\epsilon^{T}(\omega)\label{eq:7200-300-1-1-1-1-1}
\end{equation}
This media can not be realized in empty space.

\subsection{Corresponding to reverse reciprocity theorem}

\[
\intop_{S}(E_{1}(\omega)\times H_{2}^{*}(\omega)-E_{2}^{*}(\omega)\times H_{1}(\omega))\:\hat{n}dS
\]
\begin{equation}
=\intop_{V}(J_{1}(\omega)\cdot E_{2}^{*}(\omega)-J_{2}^{*}(\omega)\cdot E_{1}(\omega)-K_{1}(\omega)\cdot H_{2}^{*}(\omega)+K_{2}^{*}(\omega)\cdot H_{1}(\omega))\, dV\label{eq:7400-50-1}
\end{equation}
\begin{equation}
\epsilon_{1}(\omega)=-\epsilon_{2}^{\dagger}(\omega),\ \ \ \ \ \ \ \ \mu_{1}(\omega)=-\mu_{2}^{\dagger}(\omega)\label{eq:7400-10-1-1}
\end{equation}

\[
\intop_{S}(E_{1}(\omega)\times\frac{1}{Z}E_{2}^{*}(\omega)-ZH_{2}^{*}(\omega)\times H_{1}(\omega))\:\hat{n}dS
\]
\begin{equation}
=\intop_{V}(J_{1}(\omega)\cdot ZH_{2}^{*}(\omega)+\frac{1}{Z}K_{2}^{*}(\omega)\cdot E_{1}(\omega)-K_{1}(\omega)\cdot\frac{1}{Z}E_{2}^{*}(\omega)-ZJ_{2}^{*}(\omega)\cdot H_{1}(\omega))\, dV\label{eq:7400-50-1-1}
\end{equation}

\[
\intop_{S}(E_{1}(\omega)\times E_{2}^{*}(\omega)-Z^{2}H_{2}^{*}(\omega)\times H_{1}(\omega))\:\hat{n}dS
\]
\begin{equation}
=\intop_{V}(Z^{2}J_{1}(\omega)\cdot H_{2}^{*}(\omega)+K_{2}^{*}(\omega)\cdot E_{1}(\omega)-K_{1}(\omega)\cdot E_{2}^{*}(\omega)-Z^{2}J_{2}^{*}(\omega)\cdot H_{1}(\omega))\, dV\label{eq:7400-50-1-1-1}
\end{equation}
\begin{equation}
\epsilon_{1}(\omega)=\frac{1}{Z^{2}}\mu_{2}^{\dagger}(\omega),\ \ \ \ \ \ \ \ \mu_{1}(\omega)=Z^{2}\epsilon_{2}^{\dagger}(\omega)\label{eq:7400-10-1-1-1}
\end{equation}
if the word ``modified'' is taken away, there is
\begin{equation}
\epsilon(\omega)=\frac{1}{Z^{2}}\mu^{\dagger}(\omega),\ \ \ \ \ \ \ \ \mu(\omega)=Z^{2}\epsilon^{\dagger}(\omega)\label{eq:7400-10-1-1-1-1}
\end{equation}
This media can be realized in empty space.

\subsection{Corresponding to reverse mutual energy theorem}

\[
-\intop_{S}(E_{1}(\omega)\times H_{2}(\omega)+E_{2}(\omega)\times H_{1}(\omega))\:\hat{n}dS
\]
\begin{equation}
=\intop_{V}(J_{1}(\omega)\cdot E_{2}(\omega)+E_{1}(\omega)\cdot J_{2}(\omega)+K_{1}(\omega)\cdot H_{2}(\omega)+H_{1}(\omega)\cdot K_{2}(\omega))\, dV\label{eq:7200-250-2}
\end{equation}

\begin{equation}
\epsilon_{1}(\omega)=-\epsilon_{2}^{T}(\omega),\ \ \ \ \ \ \ \ \ \mu_{1}(\omega)=-\mu_{2}^{T}(\omega)\label{eq:7200-300-2-1}
\end{equation}

\[
-\intop_{S}(E_{1}(\omega)\times(\frac{1}{Z}E_{2}(\omega))+ZH_{2}(\omega)\times H_{1}(\omega))\:\hat{n}dS
\]
\begin{equation}
=\intop_{V}(J_{1}(\omega)\cdot ZH_{2}(\omega)+E_{1}(\omega)\cdot(-\frac{1}{Z}K_{2}(\omega))+K_{1}(\omega)\cdot\frac{1}{Z}E_{2}(\omega)+H_{1}(\omega)\cdot(-ZJ_{2}(\omega))\, dV\label{eq:7200-250-2-2}
\end{equation}
after the transform it becomes,
\[
-\intop_{S}(E_{1}(\omega)\times E_{2}(\omega)+Z^{2}H_{2}(\omega)\times H_{1}(\omega))\:\hat{n}dS
\]
\begin{equation}
=\intop_{V}(Z^{2}J_{1}(\omega)\cdot H_{2}(\omega)-E_{1}(\omega)\cdot K_{2}(\omega)+K_{1}(\omega)\cdot E_{2}(\omega)-Z^{2}H_{1}(\omega)\cdot J_{2}(\omega))\, dV\label{eq:7200-250-2-2-1}
\end{equation}

\begin{equation}
\epsilon_{1}(\omega)=\frac{1}{Z^{2}}\mu_{2}^{T}(\omega),\ \ \ \ \ \ \ \ \ \mu_{1}(\omega)=Z^{2}\epsilon_{2}^{T}(\omega)\label{eq:7200-300-2-1-1}
\end{equation}
if the ``modified'' is taken away, there is
\begin{equation}
\epsilon(\omega)=\frac{1}{Z^{2}}\mu^{T}(\omega),\ \ \ \ \ \ \ \ \ \mu(\omega)=Z^{2}\epsilon^{T}(\omega)\label{eq:7200-300-2-1-1-1}
\end{equation}
This kind of media can be realized in empty space.

\subsection{Corresponding to mutual energy theorem}

\[
-\intop_{S}(E_{1}(\omega)\times H_{2}^{*}(\omega)+E_{2}^{*}(\omega)\times H_{1}(\omega))\:\hat{n}dS
\]
\begin{equation}
=\intop_{V}(J_{1}(\omega)\cdot E_{2}^{*}(\omega)+E_{1}(\omega)\cdot J_{2}^{*}(\omega)+K_{1}(\omega)\cdot H_{2}^{*}(\omega)+H_{1}(\omega)\cdot K_{2}^{*}(\omega))\, dV\label{eq:7200-250-2-1}
\end{equation}
\begin{equation}
\epsilon_{2}(\omega)=\epsilon_{1}^{\dagger}(\omega),\ \ \ \ \ \ \ \ \ \mu_{2}(\omega)=\mu_{1}^{\dagger}(\omega)\label{eq:7200-300-1-2-2}
\end{equation}
after transform, it becomes

\[
\zeta_{s}=s\zeta=[ZH,\frac{1}{Z}E,-\frac{1}{Z}K,-ZJ,-\frac{1}{Z^{2}}\mu,-Z^{2}\epsilon]
\]

\[
-\intop_{S}(E_{1}(\omega)\times\frac{1}{Z}E_{2}^{*}(\omega)+ZH_{2}^{*}(\omega)\times H_{1}(\omega))\:\hat{n}dS
\]
\begin{equation}
=\intop_{V}(J_{1}(\omega)\cdot ZH_{2}^{*}(\omega)+E_{1}(\omega)\cdot(-\frac{1}{Z})K_{2}^{*}(\omega)+K_{1}(\omega)\cdot\frac{1}{Z}E_{2}^{*}(\omega)+H_{1}(\omega)\cdot(-Z)J_{2}^{*}(\omega))\, dV\label{eq:7200-250-2-1-2}
\end{equation}
or

\[
-\intop_{S}(E_{1}(\omega)\times E_{2}^{*}(\omega)+Z^{2}H_{2}^{*}(\omega)\times H_{1}(\omega))\:\hat{n}dS
\]
\begin{equation}
=\intop_{V}(Z^{2}J_{1}(\omega)\cdot H_{2}^{*}(\omega)-E_{1}(\omega)\cdot K_{2}^{*}(\omega)+K_{1}(\omega)\cdot E_{2}^{*}(\omega)-Z^{2}H_{1}(\omega)\cdot J_{2}^{*}(\omega))\, dV\label{eq:7200-250-2-1-2-1}
\end{equation}

\begin{equation}
\epsilon_{1}(\omega)=-\frac{1}{Z^{2}}\mu_{2}^{\dagger}(\omega),\ \ \ \ \ \ \ \ \ \mu_{1}(\omega)=-Z^{2}\epsilon_{2}^{\dagger}\label{eq:7200-300-1-2-2-1}
\end{equation}
If the ``modified'' is taken away, there is

\begin{equation}
\epsilon(\omega)=-\frac{1}{Z^{2}}\mu^{\dagger}(\omega),\ \ \ \ \ \ \ \ \ \mu(\omega)=-Z^{2}\epsilon^{\dagger}\label{eq:7200-300-1-2-2-1-1}
\end{equation}
This media can not realized in empty space.

If inverse Fourier transform is made, we can obtained the 4 corresponding
time domain mutual energy or reciprocity theorems.

\section{Conclusion}

The modified Poynting theorem is introduced, so it functions for the
superimposition of the electromagnetic fields which includes retarded
potential, mirrored field of the retarded potential which is a field
of advanced potential and time-offset electromagnetic fields. Each
fields can also be in a different media.

The concept of the mutual energy is introduced, which is the difference
between the total energy and self-energy. Using the concept of the
mutual energy a few mutual energy theorems are derived from the modified
Poynting theorem. The mutual energy theorems introduced in this article
includes, 

1) The instantaneous-time mutual energy theorem. 

2) The time-reversed mutual energy theorem.

3) Mixed mutual energy theorem. 

4) The time-correlation reciprocity theorem is re-derived from the
above instantaneous-time mutual energy theorem. Hence it can be referred
as time-correlation mutual energy theorem too. 

5) The mutual energy theorem in Fourier domain is re-derived from
Poynting theorem in Fourier domain and time domain 

6) The Lorenz reciprocity theorem is re-derived as a special case
of the mutual energy theorem, where the electromagnetic fields are
opposite, one is retarded potential and the other is mirrored field
of retarded potential which is advanced potential. The concept of
the reaction is also explained as a special mutual energy where two
field one is retarded potential one is advanced potential. 

7) We also extended the mutual energy theorem to the case there are
$N$ electromagnetic fields instead of only 2. Since the use of inner
product, the formula for $N$ electromagnetic fields is very simple
and easy to understand. 

8) the 3 additional complementary theorems are derived.

This article has built a bridge between the Poynting theorem and the
reciprocity theorem. This bridge is mutual energy theorem.

\section*{Appendix 1}

Assume $f(t)$ and $g(t)$ is real function, and $f(\omega)=F\{f(t)\}$,
$g(\omega)=F\{g(t)\}$

\begin{equation}
F\{\intop_{t=-\infty}^{\infty}f(t+\tau)\, g(t)dt\}=f(\omega)(g(\omega))^{*}\label{eq:A-1-10}
\end{equation}

\section*{Appendix 2}

Prove the formula

\begin{equation}
\intop_{t=-\infty}^{\infty}(\xi_{2}(t),\partial_{t+\tau}\eta_{1}(t+\tau))+(\xi_{1}(t+\tau),\partial_{t}\eta_{2}(t))\, dt=0\label{eq:5000-10}
\end{equation}
In case there is 
\begin{equation}
\epsilon_{2}^{\dagger}(\omega)-\epsilon_{1}(\omega)=0,\ \ \ \ \ \ \mu_{2}^{\dagger}(\omega)-\mu_{1}(\omega)=0\label{eq:5000-11}
\end{equation}

Do the Fourier transform $F\{\bullet\}$ to the above formula,

\[
F\{\intop_{t=-\infty}^{\infty}(\xi_{2}(t),\partial_{t+\tau}\eta_{1}(t+\tau))+(\xi_{1}(t+\tau),\partial\eta_{2}(t))dt\}
\]
\[
=F\{\intop_{V}\intop_{t=-\infty}^{\infty}(E_{2}(t)\cdot\partial_{t+\tau}D_{1}(t+\tau)+E_{1}(t+\tau)\cdot\partial D_{2}(t)
\]
\begin{equation}
+H_{2}(t)\cdot\partial_{t+\tau}B_{1}(t+\tau)+H_{1}(t+\tau)\cdot\partial B_{2}(t))\, dt\, dV\}\label{eq:5000-20}
\end{equation}
Let$U=\partial_{t}D_{1}(t)$
\[
F\{\intop_{t=-\infty}^{\infty}(E_{2}(t)\cdot\partial_{t+\tau}D_{1}(t+\tau))\, dt\}
\]
\[
=F\{\intop_{t=-\infty}^{\infty}(E_{2}(t)\cdot U(t+\tau))\, dt\}
\]
\begin{equation}
=E_{2}^{*}(\omega)U(\omega)\label{eq:5000-30}
\end{equation}
Where
\begin{equation}
U(\omega)=F\{U\}=F\{\partial_{t}D_{1}(t)\}=\{\partial_{t}\intop_{\tau=-\infty}^{\infty}\epsilon_{1}(t-\tau)E_{1}(\tau)d\tau\}=-j\omega\epsilon_{1}(\omega)E_{1}(\omega)\label{eq:5000-40}
\end{equation}
or

\begin{equation}
U^{*}(\omega)=(-j\omega)^{*}\epsilon_{2}^{*}(\omega)E_{2}^{*}(\omega)\label{eq:5000-50}
\end{equation}
Hence the Eq.(\ref{eq:5000-20}) becomes
\[
=\intop_{V}(E_{2}^{*}(\omega)\cdot(-j\omega)\epsilon_{1}(\omega)E_{1}(\omega)+E_{1}(\omega)\cdot(-j\omega)^{*}\epsilon_{2}^{*}(\omega)E_{2}^{*}(\omega)
\]
\[
+H_{2}^{*}(\omega)\cdot(-j\omega)\mu_{1}(\omega)H_{1}(\omega)+H_{1}(\omega)\cdot(-j\omega)^{*}\epsilon_{2}^{*}(\omega)H_{2}^{*}(\omega))dV
\]
\[
=(-j\omega)\intop_{V}(E_{2}^{*}(\omega)(\epsilon_{1}(\omega)-\epsilon_{2}^{\dagger}(\omega))E_{1}(\omega)dV
\]
\[
+(-j\omega)\intop_{V}(H_{2}^{*}(\omega)(\mu_{1}(\omega)-\mu_{2}^{\dagger}(\omega))H_{1}(\omega)dV
\]
\[
=0
\]
Where we have considered that

\begin{equation}
E_{1}\epsilon_{2}^{*}E_{2}^{*}=E_{2}^{*}(\epsilon_{2}^{*})^{T}E_{1}=E_{2}^{*}\epsilon_{2}^{\dagger}E_{1}\label{eq:5000-70}
\end{equation}
\begin{equation}
E_{1}\mu_{2}^{*}E_{2}^{*}=E_{2}^{*}(\epsilon_{2}^{*})^{T}E_{1}=E_{2}^{*}\epsilon_{2}^{\dagger}E_{1}\label{eq:5000-70-1}
\end{equation}
The last step Eq.(\ref{eq:5000-11}) has been considered. Hence we
have Eq.(\ref{eq:5000-10}),

\begin{equation}
\epsilon_{2}^{\dagger}=\epsilon_{1}\label{eq:5000-80}
\end{equation}
\begin{equation}
\mu_{2}^{\dagger}=\mu_{1}\label{eq:5000-80-1}
\end{equation}

\section*{Appendix 3}

Prove if $\xi_{1}$is retarded potential and $\xi_{2}$ is advanced
potential and the integral satisfies that,

\begin{equation}
(\xi_{1},\xi_{2})=0\label{eq:450-5}
\end{equation}
We can assume $\xi_{2}=h\,\xi_{2o}$, here $h$ is magnetic mirror
transform. $\xi_{2o}$ is the corresponding field of $\xi_{2}$. $\xi_{2o}$
is retarded potential, 

\[
(\xi_{1},\xi_{2})=\intop_{S}(E_{1}(t)\times H_{2}(t)+E_{2}(t)\times H_{1}(t))\,\hat{n}dS
\]
\[
=\intop_{S}(E_{1}(t)\times(-1)H_{2o}(-t)+E_{2o}(-t)\times H_{1}(t))\,\hat{n}dS
\]
\begin{equation}
=(-\intop_{S}(E_{1}(t)\times H_{2o}(\tau-t)-E_{2o}(\tau-t)\times H_{1}(t))\,\hat{n}dS]_{\tau=0}\label{eq:A-3-10}
\end{equation}
In order to the above formula, the Fourier transform is applied, 

\begin{equation}
f(\tau)\equiv\intop_{S}(E_{1}(t)\times H_{2o}(\tau-t)-E_{2o}(\tau-t)\times H_{1}(t))\,\hat{n}dS\label{eq:A-3-20}
\end{equation}
\begin{equation}
F\{f(\tau)\}=\intop_{S}(E_{1}(\omega)\times H_{2o}(\omega)-E_{2o}(\omega)\times H_{1}(\omega))\,\hat{n}dS\label{eq:A-3-30}
\end{equation}
Where $F\{\bullet\}$ is Fourier transform. In the big sphere. Assume
$r\rightarrow\infty$.Considering the Silver-Muller radiation condition
or Sommerfeld's radiation condition,
\begin{equation}
\lim_{r\rightarrow\infty}r(H\times\hat{n}-E)=0\label{eq:A-3-40}
\end{equation}
In $r\rightarrow\infty$, $\hat{n}$ can be calculated, 

\begin{equation}
\hat{n}=\frac{E_{1}(\omega)\times H_{1}(\omega)}{||E_{1}(\omega)\times H_{1}(\omega)||}\label{eq:A-3-50}
\end{equation}
and
\begin{equation}
\hat{n}=\frac{E_{2o}(\omega)\times H_{2o}(\omega)}{||E_{2o}(\omega)\times H_{2o}(\omega)||}\label{eq:A-3-60}
\end{equation}
or at $r\rightarrow\infty$ there is 
\begin{equation}
E_{1}(\omega)=E_{1}(\omega)\times\hat{n}\label{eq:A-3-70}
\end{equation}
\begin{equation}
E_{2o}(\omega)=E_{2o}(\omega)\times\hat{n}\label{eq:A-7-80}
\end{equation}

\[
F\{f(\tau)\}=\intop_{S}(E_{1}(\omega)\times(E_{2o}(\omega)\times\hat{n})-E_{2o}(\omega)\times(E_{1}(\omega)\times\hat{n}))\,\hat{n}dS
\]
\[
=\intop_{S}(E_{1}(\omega)\cdot E_{2o}(\omega))\hat{n}-(E_{2o}(\omega)\cdot E_{1}(\omega))\hat{n})\,\cdot\hat{n}dS
\]
\begin{equation}
=0\label{eq:A-3-90}
\end{equation}
In the above we have considered
\begin{equation}
a\times(b\times c)=(a\cdot b)c-(a\cdot c)b\label{eq:A-3-100}
\end{equation}
That means that 

\begin{equation}
F\{f(\tau)\}=0\label{eq:A-3-110}
\end{equation}
Hence

\begin{equation}
f(\tau)=0\label{eq:A-3-120}
\end{equation}
i.e.
\begin{equation}
f(\tau)=\intop_{S}(E_{1}(t)\times H_{2o}(\tau-t)-E_{2o}(\tau-t)\times H_{1}(t))\,\hat{n}dS=0\label{eq:A-3-130}
\end{equation}
Hence
\begin{equation}
\intop_{S}(E_{1}(t)\times H_{2o}(\tau-t)-E_{2o}(\tau-t)\times H_{1}(t))\,\hat{n}dS)_{\tau=0}=0\label{eq:A-3-140}
\end{equation}
or
\begin{equation}
(\intop_{S}(E_{1}(t)\times H_{2o}(-t)-E_{2o}(-t)\times H_{1}(t))\,\hat{n}dS=0\label{eq:A-3-150}
\end{equation}
or

\[
(\xi_{1},\xi_{2})
\]
\begin{equation}
=\intop_{S}(E_{1}(t)\times H_{2o}(-t)-E_{2o}(-t)\times H_{1}(t))\,\hat{n}dS=0\label{eq:A-3-160}
\end{equation}

\end{document}